\documentclass[iop,revtex4]{emulateapj}
\usepackage{natbib}
\usepackage{epsfig}
\usepackage{lscape}
\usepackage{hyperref} 
\usepackage{graphicx}

\begin{document}

\DeclareGraphicsExtensions{.pdf,.gif,.jpg,.eps,.ps}

\title{Herschel Survey of the Palomar-Green QSOs at Low Redshift}
%\author{Petric et al. }
\author{Andreea O.  Petric\altaffilmark{1}, Luis C. Ho\altaffilmark{2,3}, Nicolas J. M. Flagey\altaffilmark{4} , Nicholas Z. Scoville\altaffilmark{5}}

\altaffiltext{1}{Gemini Observatories, 670 N. A'Ohoku Pl., Hilo, HI, 96720,USA}
\altaffiltext{2}{Kavli Institute for Astronomy and Astrophysics, Peking University, Beijing 100871, China}	
\altaffiltext{3}{Department of Astronomy, School of Physics, Peking University, Beijing 100871, China}
\altaffiltext{4}{Institute for Astronomy, 640 North AÔOhoku Place, Hilo, HI 96720,USA}
\altaffiltext{5}{California Institute of Technology, 1200 E. California Blvd. , MS 249-17, 91125,USA}

%\altaffiltext{3}{Jet Propulsion Laboratory}
\begin{abstract}
We investigate the global cold dust properties of 85 nearby ($z \leq 0.5$) QSOs, chosen from the Palomar-Green sample of optically luminous quasars. We determine their infrared spectral energy distributions and estimate their rest-frame luminosities by combining {\it{Herschel}} data from 70 to 500 $\mu$m with near-infrared and mid-infrared measurements from the Two Micron All Sky Survey (2MASS) and the Wide-Field Infrared Survey Explorer (WISE). In most sources the far-infrared (FIR) emission can be attributed to thermally heated dust. Single temperature modified black body fits to the FIR photometry give an average dust temperature for the sample of 33~K, with a standard deviation of 8~K, and an average dust mass of $7 \times 10^6 M_{\odot}$ with a standard deviation of $9 \times 10^6 M_{\odot}$. Estimates of star-formation that are based on the FIR continuum emission correlate with those based on the 11.3~$\mu$m PAH feature, however, the star-formation rates estimated from the FIR continuum are higher than those estimated from the 11.3 $\mu$m PAH emission. We attribute this result to a variety of factors including the possible destruction of the PAHs and that, in some sources, a fraction of the FIR originates from dust heated by the active galactic nucleus and by old stars. 

\end{abstract}

\section{Introduction}
Insight into the properties and evolution of the interstellar medium (ISM) is critical for building theories of how galaxies evolve. The ISM is the birthplace of stars, is enriched by them, and feeds super-massive black holes (BH). Galaxies evolve by transforming their gas and dust reservoirs into stars at low rates, through occasional dramatic interactions, and by capturing gas and stars from the intergalactic medium. 

The consequence of those events appears to be that the masses of central BH are correlated with the stellar mass and velocity dispersion of the bulge. This was suggested by observations of the dynamics of stars and gas in the nuclear regions of nearby galaxies \citep[e.g.][and references therein]{magor1998, FerMer2000,Geb2000,Trem2002, kho2013}. 

A frequently offered explanation for this correlation is that the bulge and the BH control each other's growth through their interactions with the ISM. AGN winds and/or jets impact the gas and dust in the surrounding ISM (i.e. feedback). The AGN radiation can heat, ionize  and/or push out the ISM; this quenches star-formation and limits the amount of fuel available for accretion into the BH. AGN related activity such as powerful radio jets can also have the opposite effect, that is, jets drive shocks through the ISM, create regions of compressed gas facilitating enhanced star-formation. \citet{kho2013} show that differences between the light profiles of ellipticals can be related to the type of AGN feedback they experienced. To test models of how AGN affect the ISM we need an accurate census of the ISM in AGN hosts as a function of AGN properties. 

One practical method to estimate the mass of cold ISM is to measure the total dust mass and then use an appropriate gas-to-dust ratio to infer the total cold ISM. In this paper we focus on the first step: measure the FIR emission and assess whether it is thermal dust emission in star-forming regions. 

Observations in the FIR of large samples of nearby, luminous, broad-line QSOs were done with the Infrared Astronomical Satellite (IRAS) by \citet{neug1986} and \citet{sand1988a}. \citet{sand1988a,sand1988b} suggested that QSOs were preceded by a dusty ULIRG phase. \citet{haas2003} observed a sample of 64 PG QSOs with the {\it{Infrared Space Observatory}} (ISO) satellite and suggested that the observed diversity of SEDs implied changes in the dust distribution as the host galaxies evolved.  Pre-ALMA studies of molecular and neutral gas emission \citep{evans2001,scov2003,bert2007,ho2008a} made it clear that a significant fraction of optically selected quasars have considerable amounts ($\sim 10^9-10^{10} \rm{M}_{\odot}$) of cold gas.  

With unprecedented high resolution and sensitivity, the European Space Agency's {\it{Herschel Space Observatory}} \citep{pil2010}  dramatically increased our ability to measure the ISM of nearby and higher redshift sources \citep[e.g.][]{lutz2014}. Here we present measurements of cold dust in a sample of 85 of the 87 nearby ($z \leq 0.5$), optically luminous, broad-line QSOs selected from the Palomar Bright Quasar Survey Catalogue of \citet{sgreen1983} with redshifts from \citet{bgreen1992} \footnote{We only present FIR photometry for 85 out of the 87 sources in the \citep{bgreen1992} sample because we shared several targets with other Herschel programs, which were not completed.}

The QSOs targeted in this FIR survey have BH masses in the range of known BH masses ($M_{BH}$) from $\sim 10^{6.5}$ to $10^{9.5}$ M$_{\odot}$ and $L_{bol}/L_{Edd} $ from 0.03 to 1 \citep{boro2002}. Multi-wavelength studies from X-ray to radio \citep[e.g][]{elvis1986, bech1987, kel1989, evans2001, scov2003, guy2006, dasyra2007, vei2009} help place the Herschel observations in a wider astrophysical context.

We compare FIR observations to radio and MIR data to investigate the origin of the FIR. We use the radio observations to estimate how much of the observed FIR could be synchrotron radiation in addition to thermal dust emission. %It has been shown that FIR and radio emission from radio-loud QSOs are highly correlated but those from radio-quite quasars appear not to be \citep{kal2014}. 
 
 We also compare the FIR continuum to PAH emission since both trace star-formation. The FIR emission is mostly large grains ($>0.1 \mu$m ) that absorb and re-radiate starlight emission and PAH emission originates from organic compounds excited by optical/UV radiation from young stars. PAH emission is weaker in local AGN than in star-forming regions \citep{roche1991} but PAH studies of AGN samples disagree about the reliability of SFR estimates from PAHs. \citet{lam2012} finds that the luminosity of the 11.3 $\mu$m feature is significantly suppressed relative to other PAH features in AGN dominated systems, while \citet{Dstan2010}, \citet{DSan2010} and \citet{esq2014} find that the 11.3 $\mu$m PAH feature can survive within 100 pc form the AGN and that for nearby Seyferts the 11.3 $\mu$m PAH is not suppressed.

 The paper is structured as follows: in section 2 we describe the observational setup and data reduction, in section 3 we present basic photometric results and basic derivations of rest frame luminosities, spectral indices and global dust masses and temperatures.  In section 4 we analyze the dust masses and temperatures relative to the radio and PAH 11.3 $\mu$m properties and in section 5 we summarize our findings and conclusions.

\section{Observations and Data Analysis}
The photometric data presented here were taken with the {\it{Herschel Space Observatory}} \citep{pil2010}. Our observations employed the Photodetector Array Camera and Spectrometer  \citep[PACS,][]{pog2010} and the Spectral and Photometric Imaging Receiver \citep[SPIRE,][]{grif2010} instruments. With PACS we measured the 70, 100, and 160 $\mu$m broadband emission and with SPIRE the 250, 350 and 500 $\mu$m emission. 

\subsection{PACS}
The PACS observations were performed in the {\it{mini-scan mode}} with scan angles of 70 and 110 degrees, as recommended by the instrument team. Our scanning speed was 20$" $/sec. PACS is a dual photometer allowing simultaneous short wavelength (70 or 100 $\mu$m) and long wavelength (160 $\mu$m) observations over a field of view of  1.75$' ~\times ~ 3.5'$. The FWHMs of the PSF are 5.46'' $\times$ 5.76'', 6.69'' $\times$ 6.89'', and 10.65'' $\times$ 12.13'' \footnote{See: {\url{http://herschel.esac.esa.int/twiki/pub/Public/PacsCalibrationWeb/bolopsf_20.pdf}}}. The integration time for each scan angle was 180 seconds. 

We processed all the PACS data using the pipeline in the Herschel Interactive Processing Environment (HIPE) and associated calibrations version 48 and above. Details of the HIPE pipeline and calibration tables are presented in the online PACS data reduction manual\footnote{ \url{http://herschel.esac.esa.int/Data_Processing.shtml }}.

We combine the two PACS scans into a final image employing standard North - America Herschel Science Center (NHSC) data reprocessing tools combined in a script provided by the NHSC PACS team. The pixel sizes of the 70, 100 and 160 $\mu$m maps are 1.2, 1.4, and 2.1". These pixel sizes were used by the PACS team in maps of calibration sources to adequately sample the PSF. 

The PACS fluxes presented here are computed by summing all pixels within circular $\sim$20" apertures centered at the centroids of the 70, 100, and 160 $\mu$m emission, respectively. For some targets we adjusted slightly the aperture size, if there was another source or a bad pixel near the source. The background was estimated from the median value of apertures placed throughout the map, avoiding positions which would include the source. We fit the distribution of fluxes in those apertures with a gaussian plus a second order polynomial. The second-order polynomial is needed to model the higher noise at the map edges. We choose the median value of the distribution of background apertures fluxes as our final background estimate noting that the median value is within a few percent of the mean value for most maps.

We estimate flux measurements errors ($\sigma$) from the standard deviation of the gaussian used to fit to the distribution of background aperture fluxes.  Upper limits are 3 $\sigma$. We  then apply the aperture correction factors provided in the official PACS documentation. The aperture corrections range between 1.1 and 1.3, 1.1 and 1.3, and 1.1 and 1.2 at 70 $\mu$m, 100 $\mu$m, and 160 $\mu$m respectively. The flux calibration accuracy of PACS observation is 5\%. 

\subsection{SPIRE}
The SPIRE observations were taken in {\it{small-scan-map mode}}. SPIRE has an effective field of view of 4$' ~\times ~ 8' $. The PSF FWHMs for the SPIRE data are 18.1", 25.2",  and 36.6" respectively. The SPIRE maps have pixel sizes of 6", 8", and 12" at 250, 350, and 500 $\mu$m, respectively. The duration of each on-source integration was 37 seconds. 

The SPIRE fluxes are estimated following the suggestions from the SPIRE instrument team \citep{pear2013}. The Herschel/SPIRE photometer's most basic data products are timeline data \citep{grif2010}. These data contain information about each of the bolometers including their positions, and voltage registered at each position as they scan a source\footnote{\url{http://herschel.esac.esa.int/twiki/pub/Public/DataProcessingWorkshop2012/Introduction_To_SPIRE_Pipeline_v1.0.pdf}}. The SPIRE-ICC (Instrument Control Center) and both the North - America and the ESA Herschel Science Centers recommend fitting the timeline data to estimate the fluxes of unresolved sources\footnote{(see SPIRE ObserverÕs Manual Section 5.2.11)}. 
  
The PSF fitter routine provided in HIPE fits a Gaussian to the signal versus position timeline data. For unresolved sources, the peak of the fitted Gaussian corresponds to the flux density of the source. This procedure is also employed to derive the flux calibration parameters for SPIRE observations \citet{bendo2013}. 
 
The SPIRE Timeline fitter uses a Levenberg-Marquardt algorithm to fit all the timeline data associated with one source and avoid smearing effects due to pixelization associated with the SPIRE map-making process. The timeline fitter requires default values for how far from a given source position it should search for a flux peak, and also what annulus outside this region it can use to estimate the background. The default values for those parameters were derived from observations of Neptune and Gamma Draconis \citep{bendo2013} and tested with observations of standard calibration stars. 

The uncertainties in the fluxes we present at 250, 350 and 500 $\mu$m are the errors from the timeline fitter. SPIRE flux calibration uncertainties are estimated to be $\sim$7\% \citep{bendo2013}. For non-detections we calculate upper limits as 3 times the standard deviation in the timeline data. The SPIRE errors are consistent with the confusion limit fluxes estimated by \citet{ngu2010}.

We test visually that the sources we deem as detections using the timeline fitter are not affected by the presence of other objects by producing maps and comparing the SPIRE sources visible in these SPIRE maps with sources that are visible  in the higher resolution PACS maps. 
 
\section{Results}
\subsection{Measurements}
Flux measurements and associated basic statistics are given in Tables \ref{Pphot} and \ref{PhotStats}. The {\it{Herschel}} data have more than an order of magnitude higher spatial resolution than previous IRAS observations and a factor of 8 better than those from ISO and we are able to detect sources that are about 4 times fainter than the faintest PG QSOs detected by IRAS and ISO. Figure \ref{FLnLUMS} shows histograms of the measured fluxes at 70, 100, 160, 250, 350, and 500 $\mu$m.

%\begin{figure*}[!h]
%$\begin{array}{cc}
%\includegraphics[angle=0,width=0.49\linewidth]{PB_FL_histo.eps}
%\includegraphics[angle=0,width=0.49\linewidth]{PG_FL_histo.eps}\\
%\includegraphics[angle=0,width=0.49\linewidth]{PR_FL_histo.eps}
%\includegraphics[angle=0,width=0.49\linewidth]{Sb_FL_histo.eps}\\
%\includegraphics[angle=0,width=0.49\linewidth]{Sg_FL_histo.eps}
%\includegraphics[angle=0,width=0.49\linewidth]{Sr_FL_histo.eps}\\
%\end{array}$
%\caption{{\small{Histograms observed Herschel PACS and SPIRE fluxes at 70, 100, 160, 250, 350, and 500 $\mu$m. The upper limits are shown as dashed histograms. }} \label{FLnLUMS}}
%\end{figure*}

NIR and MIR from the Two Micron All Sky Survey \citep[2MASS]{skru2006} and Wide-field Infrared Survey Explorer \citep[WISE]{wrig2010} surveys are given in Table \ref{NMIRtab}. We obtained those by searching the IRSA catalogs within a radius of 15" of the optical position. This search radius was chosen to be comparable to the 12" angular resolution of the 22 $\mu$m WISE observations and include potential WISE pointing errors.  Two examples of the NIR through FIR SEDs are plotted in Figure \ref{PSEDs}. The electronic version presents the SEDs for all the 85 targets. 
 
 Most SEDs peak at $\sim100 ~\mu$m, (Figure {\ref{PSEDs}) suggesting that the observed FIR is thermal emission from cold ($\sim$ 40K ) dust. The shapes of most SEDs are modified black-bodies, therefore we fit modified black bodies to the FIR data (see section 3.4). The error bars of the NIR and MIR fluxes, are also plotted but are significantly smaller than the relative FIR errors. In appendix A2 we investigate the radio properties of the PG SOs to estimate the amount of synchrotron radiation which contributes to the observed FIR emission in radio loud PG QSOs.  Ancillary radio data at 1.4 and 5 GHz suggest that in 8\% of the sources the emission at wavelengths longer than 250 $\mu$m may contain synchrotron emission associated with radio loud AGN activity. 

%\begin{figure*}[!h]
%$\begin{array}{cc}
%\includegraphics[angle=90,width=0.49\linewidth]{PG0050+124_SED.eps}
%\includegraphics[angle=90,width=0.49\linewidth]{PG1302-102_SED.eps}\\
%\end{array}$
%\caption{NIR to FIR SEDs of PG 0050+124 and radio loud source PG 1302-102.  All the SEDs are available in the electronic version of this paper. \label{PSEDs}}
%\end{figure*}

\subsection{Comparison with previous measurements}
In this section we compare the Herschel photometry with previous measurements. 
\subsubsection{ISO 100 $\mu$m} 
Figure \ref{iso100} compares 31 PG QSOs observed with both Herschel and ISO at 100 $\mu$m. Measurements are consistent within detection sensitivities. Two sources have ISO fluxes that are much lower than the Herschel measurements. Among those, on target was also observed at 70$\mu$m with MIPS and the Herschel flux is within 5\% of the MIPS flux.  
In sum, Figure \ref{iso100} illustrates that at the bright end the ISO and Herschel fluxes agree very well and at the faint end the Hershel observations are factor of few to a magnitude more sensitive. 

%\begin{figure*}[!h]
% \includegraphics[angle=90,width=0.6\linewidth]{Comp_MIPS70.eps}\\
% \includegraphics[angle=90,width=0.6\linewidth]{Comp_ISO.eps}\\
%  \includegraphics[angle=90,width=0.6\linewidth]{Comp_shi160.eps}\\
% \caption{Herschel measurements at 70, 100 and 160 $\mu$m presented in this paper versus MIPS 70$\mu$m measurements from \citet{shi2014}, ISO 100 $\mu$m data from ISO \citet{haas2003} and MIPS (stars) and Herschel (crosses) measurements from \citet{shi2014} \label{iso100}}
%\end{figure*}

\subsubsection{Spitzer MIPS 70 $\mu$m} 
In figure \ref{iso100} we also compare our measurements to those from MIPS \citep{shi2014}. The PACS photometry presented in this paper are consistent on average with the MIPS observations, but have higher resolution (e.g. at 160$\mu$m SPITZER MIPS has a resolution of   while Herschel PACS images have a spatial resolution of   ) %but because of Herschel's enhanced sensitivity we detect sources at 50\% lower flux levels.% and are able to estimate upper limits at 10\% lower levels. 

\subsubsection{MIPS and Herschel 160 $\mu$m} 
\citet{shi2014} present Herschel PACS and Spitzer MIPS 160$\mu$m measurements for 78 of the 85 PG QSOs we present here. Figure \ref{iso100} shows those are consistent with the values in this work.  

\subsection{Rest Frame Luminosities}
 WISE observations at 5.5, 8 and 24 $\mu$m are combined with Herschel 70, 100, 160, 250, 350 and 500 $\mu$m data to estimate rest-frame luminosities at several key wavelengths, i.e. wavelengths that are often used to characterize the MIR and FIR properties of galaxy samples. This was done as follows: for each observed flux we determine the corresponding rest-frame $\lambda_{rest} ~=~ \lambda _{obs}/(1 ~+ ~z)$ and then compute the $F_{\lambda _{rest}}$ flux. We then derive spectral indices ($\alpha : F_{\lambda} \propto \lambda ^{\alpha} $) from each pair of adjacent spectral measurements. Finally we use these spectral indices to estimate the rest-frame fluxes at rest-frame 5.5, 8, 24, 60, 70, 100, 160, 250, 350, and 500 $\mu$m.
%\begin{figure*}[!h]
%\includegraphics[angle=90,width=0.49\linewidth]{L60_r.eps}
%\includegraphics[angle=90,width=0.49\linewidth]{L100_r.eps}\\
%\includegraphics[angle=90,width=0.49\linewidth]{L160_r.eps}
%\includegraphics[angle=90,width=0.49\linewidth]{L250_r.eps}\\
%\includegraphics[angle=90,width=0.49\linewidth]{L350_r.eps}
%\includegraphics[angle=90,width=0.49\linewidth]{L500_r.eps}\\
%\caption{Histograms of rest-frame luminosities at 60-500 $\mu$m obtained from {\it{Herschel}} photometric measurements. The upper limits are shown as dashed histograms.  \label{Rlums}}
%\end{figure*}

We list rest-frame luminosities emitted at 5.5, 8, 24, 60, 70, 100, 160, 250, 350, and 500 $\mu$m in Table {\ref{LumRestTab}}, and show histograms of rest frame luminosities at 60, 100, 160, 250, 350 and 500 $\mu$m in Figure {\ref{Rlums}}. We present basic statistics in Table \ref{LirStats}: minimum, maximum, median luminosities,  and the medians of the distributions including the upper limits derived using the Kaplan Meier estimators\footnote{ as implemented in the statistics package $\it{R}$  \citep{FB2012}}. The PG QSOs span a wide range of IR luminosities from values comparable to the IR luminosities of normal galaxies to those of Ultraluminous Infrared Galaxies (ULIRGs), which is consistent with previous findings (e.g. \citep{haas2003}) 

In Table {\ref{LumRestTab}} we also present FIR luminosities, from 40 to 500 $\mu$m and the 42.5 to 122.5 $\mu$m FIR \citep[e.g.][]{hel1985,condon1992}:
\begin{equation}
\left( {{FIR \over{W m^{-2}}}}\right) ~= ~ 1.26 ~\times ~ 10^{-14} \left( {{2.58 S_{60 \mu m} ~+~ S_{100 \mu m}}\over{Jy}}\right)
\label{CIR1}
\end{equation}
where $S_{60 \mu m} ~\rm{and} ~S_{100 \mu m}$ are the flux densities in rest frame at 60 and 100 $\mu$m. 

\subsection{SED templates}
%\begin{figure*}[!h]
%\includegraphics[angle=90,width=0.9\linewidth]{Template5SU.eps}\\
%\caption{The combined photometry of all the 85 sources in our sample normalized by the estimated luminosity at 5.5 $\mu$m. Top: solid line gives the median template computed in each observed photometric band using the Kaplan-Meier estimator, the dotted lines show the 95\% confidence interval. \label{Temps}}
%\end{figure*}
Composite SEDs are frequently used to compare different populations of objects, to study their evolution, and to assess whether selecting specific sources in one wavelength regime biases our understanding of their properties at other wavelengths.  \citep[e.g.][]{boy1990, cris1990, franc1991, elvis1994, zheng1997, bro2001, vberk2001, telf2002, scott2004, gli2006, han2013}. 

In this paper we present a template combining SEDs of nearby, optically luminous, broad-line QSOs, normalized at 6 $\mu$m (Figure {\ref{Temps}}). 
We combine the SEDs as follows: (1) we bin the rest-frame fluxes estimated from each observed photometric band, and then (2) we compute, the median, and the 95\% confidence intervals using the Kaplan-Meier estimator. We present those in table \ref{TempNorm}. 
We speculate that the heterogeneity in observed FIR properties may be connected to a variety of evolutionary paths for these sources and that the hosts of galaxies evolve slower than the central accreting SMBH, because the emission most closely associated with the AGN and the torus (i.e. optical and NIR fluxes) are fairly homogenous. 

Substantial effort has recently been put in using rest-frame far-infrared emission as a tracer of star-formation in high redshift AGN. This relies on the FIR in those systems typically not being strongly contaminated by AGN heated emission. One way to support this semi-quantitatively is by comparison to intrinsic AGN SEDs typically derived by decomposing AGN SEDs and host with the aid of Spitzer spectroscopy. The most common AGN SEDs of this type are by \citet{net2007} and \citet{mul2011}. In Figure \ref{TempsC} we compare the median SEDs derived from the PG QSOs to the average observed SED of 28 PG QSOs observed with IRS from \citet{net2007} and to the average SED of 8 FIR-weak PG QSOs from which the starburst component was subtracted \citet{net2007}. We also compare the template from the data presented in this paper to the average SEDs of nearby AGN with high and low luminosities, i.e. with log(L$_{2-10 keV} \geq \& \leq 42.9$ respectively) from \citet{mul2011}. The templates from \citet{net2007} and \citet{mul2011} seem to have relatively higher MIR emission originating from the AGN and lower FIR emission than the template presented here from WISE and Herschel photometry. This difference is maintained but slightly reduced if we exclude radio loud QSOs from our sample as well as AGN with $S_{350\mu m} \geq S_{250\mu m}$. These differences suggest that the templates  \citet{net2007} and \citet{mul2011} represent sources with hotter dust, or smaller grains emitting in the MIR. As shown in Figure \ref{TempsC} starburst galaxy templates as derived by \citet{mul2011} have comparatively more large old grains emitting in the FIR. Note that both \citet{net2007} and \citet{mul2011} estimated the intrinsic AGN-only template. We leave a similar, more detailed analysis of the MIR and FIR data for a future investigation. 

\subsection{Dust Mass and Temperature Estimates }
Dust masses and temperatures are estimated from single temperature modified black-body fits to the far-IR SEDs, i.e. including all the PACS and SPIRE photometry (70-500 $\mu$m). The results are presented in Table \ref{TabR1}. It has been known for decades that there are multiple types of grain populations, with different size distributions and chemical compositions. While small grains are stochastically heated and significantly contribute to the emission at wavelengths shorter than $\sim$70~$\mu$m \citep{comp2010}, larger grains usually emit like a modified blackbody at an equilibrium temperature and dominate the emission at longer wavelengths, i.e. we only use flux measurements observed at wavelengths equal and longer than $70\mu m$. The far-IR emission can be used to derive an average effective temperature and a total mass for large dust particles, although for certain sources it is possible that even the $70\mu m$ emission comes from hotter dust associated with the AGN.

\citet{bia2013} compare different methods to derive dust masses and temperatures by fitting the far-IR SEDs with (1) a single temperature modified black body, (2) a dust emission model, i.e. assuming a combination of different grains populations, the details of these grain models being beyond the scope of this paper. \citet{bia2013} used two dust models: that of \citet{dli2007} and that of \citet{comp2011}. Both of these models have been used intensively in the analysis of Spitzer and Herschel data, from nearby to high redshift sources. \citet{bia2013} find that the three estimated dust masses are within 20\% of each other. He concludes that his investigation vindicates the use of the single temperature modified black body to estimate global cold dust properties in large samples of objects. However, he emphasizes the importance of using dust absorption coefficients consistent with the emissivity spectral indices used in the modified blackbody fits.

In this paper, we follow the recommendations of \citet{bia2013}: we use modified blackbody fits, with absorption coefficients and spectral indices derived from the dust model of \citet{comp2011} or that of \citet{dli2007}. We use both models to assess the uncertainties and systematics within our sample. We thus assume that $\kappa=\kappa_{0} {{\left({250 \mu m} \over{\lambda}\right)}^{\beta}}$ with $\kappa_{0}=5.01~\rm{cm}^2\rm{g}^{-1}$ with an emissivity spectral index $\beta~=1.91~$ inferred from the model of \citet{comp2011}, or $\kappa_{0}=4\rm{cm}^2\rm{g}^{-1}$ with an emissivity spectral index $\beta~=2.08~$ from the model of \citet{dli2007}.

For optically thin dust emission with a single temperature the dust mass is given by: 
\begin{equation}
  M_d = {{S_{\nu _{r}} \times D^2_L} \over{ \kappa_d ({\nu_r}) \times B(\nu_r, T_d)}}
\end{equation}
where $\nu _r$ is the rest-frame frequency, $D_L$ is the luminosity distance estimated assuming the standard concordance cosmology and the redshifts from \citet{bgreen1992}, and $B(\nu_r, T_d)$ is the Planck function at the dust temperature derived from the fit. For a source at redshift $z$, $S_{\nu_{r}}$ and $\kappa_d({\nu _r})$ are given, as a function of the observed fluxes $S_{\nu_{o}}$ and wavelengths $\lambda_{o}$, by:
\begin{equation}
  {S_{\nu_{r}}}~= ~{{S_{\nu_{o}}} \over {(1+z)}}
\end{equation} 
\begin{equation}
  \kappa_d  = \kappa_{0} {{\left[{250\mu m } \over{\lambda_{o}/(1~+~z)}\right]}^{\beta}}
\end{equation}

The quality of each fit is assessed by computing the $\chi ^2$ (Table \ref{TabR1}). We also visually check that there are enough data points at wavelengths shorter and longer than the peak to constrain the dust temperature as shown in (Figure \ref{exafit}).

Global dust temperatures and masses are estimated for 72 sources. The mean and median temperatures are  $\sim$33 and $\sim$35~K respectively and the mean and median masses are $\sim 7\times 10^6 M_{\odot}$ and $\sim 3\times 10^6 M_{\odot}$ respectively. The distributions of dust masses and temperatures are shown in Figure \ref{mbfr}. 

Systematic differences are found between the fits to the SEDs that use the grain properties of \citet{comp2011} and those using the emissivity spectral index and dust absorption coefficient based on \citet{dli2007}. The dust temperatures are almost systematically larger by 1-2~K for the former. Consequently, the dust masses are systematically larger by about 20-40\% when using the dust properties from \citet{dli2007}. This disagreement is slightly larger than that observed by \citet{bia2013}. Finally, the derived $\chi^2$ is 10-30\% lower when using the $\beta $ and $\kappa$ from \citet{comp2011}. We therefore choose to present the dust masses and temperatures estimated assuming the grain properties from \citet{comp2011}. The disagreement between the two comes directly from the differences in $\kappa$ and $\beta$. 

%\begin{figure*}[!h]
%\includegraphics[angle=90,width=0.85\linewidth]{Mdust_histo.eps}\\
%\includegraphics[angle=90,width=0.85\linewidth]{Tdust_histo.eps}\\
%\caption{Histogram of dust masses, temperatures estimated from single temperature modified black-body fits with an emissivity spectral index $\beta$ of 1.91 (see section 3.4). The dashed histograms shows the result of poor fits with only 2 or 3 photometric points, or 4 points with large errors. \label{mbfr}}
%\end{figure*}

The PG QSOs host galaxies appear to have dust masses and temperatures similar to those of nearby ellipticals but smaller than those of nearby star-forming galaxies (Figure \ref{MDC}). \citet{smith2012} use Herschel observations to estimate dust masses for 62 nearby early type galaxies (ETGs). The ETGs in the \citet{smith2012} sample have stellar masses between $\sim10^{10}$ and $\sim10^{11.5}~\rm{M}_{\odot}$ and are the most massive galaxies out of 322 from a volume limited sample of ETGs. We also compare our cold dust mass estimates to those of nearby star-forming galaxies which have stellar masses between from $\sim 10^{6.4}$ and $\sim10^{11.4}~\rm{M}_{\odot}$ \citet{skib2011} after dividing their value by 2.7 as suggested by \citet{bia2013}. This is to correct for the inconsistency between the $\beta$ they derive from the fitting and the $\beta$ they use to compute $\kappa _d $. Statistical tests including upper limits in our sample indicate that the dust masses of nearby galaxies (spirals and ellipticals) are somewhat different than those of PG QSOs with probability that they are the same of 11-12 \%. The Kaplan Meier derived medians of $log(M_D)$ for nearby Spirals, the PG QSO sample and elliptical galaxies are 6.8$^{+0.2}_{-0.3}$, 6.6$^{+0.2}_{-0.2}$, and 6.4$^{+0.4}_{+0.2}$. The wide ranges of stellar masses for the different samples, as well as the non-identical ways to estimate $log(M_D)$ upper-limits of the different samples make this comparison somewhat inconclusive. 

\section{Discussion}
In this section we discuss the origin of the FIR emission in PG QSOs and speculate about the star-formation properties of their host galaxies. We also discuss a subset of PG QSOs whose IR emission was thought to come only from AGN because IRAS and ISO did not detect them.

\subsection{FIR emission and star-formation rates}
We estimate star-formation rates (SFRs) from the FIR luminosities using the formulae from \citet{mur2011}. The estimated SFRs are between $\sim 0.4 ~\rm{and}~ \sim 300~\rm{M}_{\odot}/\rm{yr}$ and are correlated with the dust masses, with a 0.6 Spearman's rank correlation coefficient and a probability of $10^{-7}$ that they are not correlated (Figure \ref{SFvMD}). The coldest dust components dominate the estimates of total dust masses while smaller amounts of warm dust can be responsible for the the bulk of the observed FIR luminosities \citep{haas2003}. If we assume that the dust grains size distribution is roughly the same throughout the sample, the strong correlation between the FIR luminosities, or star-formation rates derived from the FIR luminosities, and the dust masses suggests a uniform heating efficiency throughout the sample which may mean that stars are the main sources heating the dust. However the lack of a very strong correlation (i.e. a Spearman correlation coefficient larger than 0.8) could be explained by the presence of additional dust components at other temperatures or multiple sources of heating. 

Star-formation rates estimated from the FIR emission are compared with SFRs derived from 11.3~$\mu$m PAH features using the measurements of \citet{shi2007} and the formulae of \citet{Dstan2010}. The SFRs derived from the IR continuum correlate well with those estimated from 11.3~$\mu$m PAH (Figure \ref{spVsf}). A Spearman rank test gives a correlation index of 0.9 with a $\sim 10^{-5} $ probability that these two quantities are not correlated. SFR rates from the PAH emission ($SFR_{PAH}$) are lower, on average, by a factor of 3 than those estimated from the FIR emission ($SFR_{FIR}$). Sources at the lowest and highest FIR luminosities show the largest differences between $SFR_{FIR}$ and $SFR_{PAH}$. Out of the 22 sources with FIR luminosities above 10$^{11} L_{\odot}$ only 8 have detectable emission at 11.3~$\mu$m. 

Some of the FIR emission may be processed dust heated by the AGN and associated with a clumpy torus.  The 11.3~$\mu$m PAH may also trace only a fraction of the total star-formation. One example where this may take place is in the merging luminous infrared galaxy system II Zw 96 \citet{inami2010}. In this source a region of a few kpc emits $\sim ~80\%$ of the IR emission of the entire system. In this region, X-ray, NIR and MIR spectroscopic diagnostics all fail to detect an AGN, although the 6.2~$\mu$m PAH EW is only half of the typical value measured in normal star-forming galaxies. Finally a significant contribution to the FIR may come from dust heated by old stars \citep[e.g.]{ken2012,lutz2014} which emit fewer UV/optical photons capable to excite the PAHs. 
 
The reliability of the 11.3~$\mu$m PAH feature as star-formation estimator in galaxies that host powerful AGN is uncertain. \citet{esq2014} found that the 11.3 $\mu$m PAH is not suppressed in the vicinity of low luminosity AGN while \citet{lam2012} found that the 11.3~$\mu$m PAH feature is significantly suppressed in the most AGN-dominated systems. The PAH observations of \citet{shi2007} were acquired with Spitzer IRS, and it is possible that the IRS projected slit missed some of the star-forming clumps that may significantly contribute to the FIR emission of the entire system. However, if this were true for a significant number of sources we would find a wider range of ratios of the PAH to the FIR luminosities throughout the sample. We note however that the \citet{Dstan2010} relation between the 11.3 ~$\mu$m PAH feature and SFR was derived for normal spiral galaxies. \citet{net2007} found that the ratios of 7.7 $\mu$m PAH to the FIR emission in PG QSOs are similar to those measured in ULIRGs, and that those ratios are lower for ULIRGs than they are for spiral galaxies.

We speculate that FIR weak sources may also have a lower number of UV photons to excite the PAHs. At high FIR luminosities there may be an enhancement of UV and optical AGN radiation which ionizes and/or destroys the PAHs. The observed FIR may be a combination of dust heated by young and old stars, as well as dust heated by the AGN. The estimated PG QSO's dust temperatures are similar to those of normal and star-forming galaxies which may suggest that the relation between SFR and dust masses is not sufficient to disentangle the processes and sources heating the FIR emitting dust.

\subsection{FIR weak QSOs} 
The star-formation history of the universe has been studied through many methods, including IR source counts. To disentangle between the AGN/starburst contribution to the IR luminosity, one can compare the IR slope and/or IR emission lines flux ratios to those measured for pure starbursts and pure AGN \citep[e.g.][]{armus2007,vei2009,petric2011}. 
 
Eight PG QSOs have been used to represent sources in which AGN contributes 100\% of the MIR emission because they were not detected by ISO and IRAS in the FIR\citep[e.g.]{net2007,vei2009}. In this work we find that some of these PG QSOs observed with the more sensitive Herschel have detectable cold dust. Out of these 8 PG QSO, 5 are in our sample: PG 0026+129, PG 1309+355, PG0923+201, PG 1617+175, PG 2251+113. All five were detected at 70 and 100 $\mu$m. %This suggests that a revision ought to be made to the AGN empirical diagnostics because a simple scaling of the PG QSOs MIR properties may over-estimate the AGN contribution to the IR. Such a revision would include SED fitting of those 5 QSOs to separate the contribution of the AGN to the bolometric luminosity, and is beyond the scope of this paper. 

\section{Conclusions}
We present high sensitivity observations taken with the {\it{Herschel Space Observatory}} to measure the cold dust content in a sample of 85 nearby ($z \leq 0.5$) QSOs chosen from the optically luminous broad-line Palomar Green (PG) QSOs sample. We detect 93\% of the sources in at least one Herschel band. The 70 $\mu$m detections range between 14 mJy and 2.3 Jy. We derive spectral indices and rest-frame luminosities at 5.5, 24, 60, 70, 100, 160, 250, 350 and 500 $\mu$m as well as FIR luminosities. The FIR luminosities span four orders of magnitude between 2 $\times ~ 10^{9} L_{\odot}$ and 1.65 $\times~ 10^{12} L_{\odot}$, i.e. between the FIR luminosities of normal galaxies to those of Ultra Luminous Infrared Galaxies. 

We combine {\it{Herschel}} data with near-infrared and mid-infrared measurements from the Two Micron All Sky Survey (2MASS) and the Wide-Field Infrared Survey Explorer (WISE) to determine their IR spectral energy distributions and to assess aggregate dust properties for 80\% of the sample. We also derive an SED template normalized at 5.5 $\mu$m. 

 We employ single temperature modified black-body fits using two models for the dust emissivity spectral index and dust absorption coefficients from \citep{ddb2007} and \citep{comp2011}, to estimate dust temperature with a mean, median and standard deviation or 33, 35 and 8~K, and dust masses $M_d$ with a median, mean, and standard deviation of  $3, 7 \times 10^6 M_{\odot}$ with a standard deviation of $9 \times 10^6 M_{\odot}$. 

We investigate how FIR indicators of star-formation compare to the 11.3 $\mu$m PAH line as measured by \citep{shi2007}. We find a tight correlation between the two star-forming indicators, and also between the luminosity and EW of the 11.3 $\mu$m PAH feature and $M_d$. %The dust colors as parametrized by 70 to 24, 100 to 24 and 250 to 24 rest-frame flux ratios are weakly correlated with the 11.3 $\mu$m PAH luminosity. 
The ratios between the SFR rate from the FIR to that from the PAH range between 1.5 and 22, with median and mean values of 8 and 9 respectively. We find fairly constant ratios of the 11.3 $\mu$m PAH luminosities to the FIR luminosities. However, at the highest FIR luminosities the ratio of the 11.3 $\mu$m PAH to the total FIR decreases in a non-linear fashion and no 11.3 $\mu$m PAH features were detected in sources with estimates of $T_d ~>~50 K$. We suggest a variety of factors that may have produced those results including that in some of the FIR brightest sources, the PAHs are being destroyed by the AGN.

\section*{Acknowledgements}
LCH acknowledges support by the Chinese Academy of Science through grant No. XDB09030102 (Emergence of Cosmological Structures) from the Strategic Priority Research Program and by the National Natural Science Foundation of China through grant No. 11473002. 

We thank the anonymous referee for providing valuable comments and help in improving the contents of this paper.

This publication makes use of data products from the Two Micron All Sky Survey, which is a joint project of the University of Massachusetts and the Infrared Processing and Analysis Center/California Institute of Technology, funded by the National Aeronautics and Space Administration and the National Science Foundation. 

AP would like to thank Mia U. and Maya Angelou for their lives and words "You may encounter many defeats, but you must not be defeated. In fact, it may be necessary to encounter the defeats, so you can know who you are, what you can rise from, how you can still come out of it " (M. A.).

\appendix
\section{A1: Notes on Individual Objects}
{\bf{PG 0007+106}}  The SPIRE fluxes measured in this source were higher than what would be expected from a modified black body that fits well the PACS emission (see section 3.4). We compared the centroids of the PACS and SPIRE emission regions and found them to be identical within the resolution of the data. Also the SPIRE fluxes obtained from fits to the time-line data are similar to those found with aperture photometry on the SPIRE maps. This source however is radio loud and variable, and some of the observed FIR emission may be synchrotron. 

{\bf{PG 0049+171}}  We present the measurement as a detection albeit with a large uncertainty in the given measurement because we do see a source in the map. PG 0049+171 is not detected by IRAS at 60 $\mu$m, the quoted IRAS measurement at 60 $\mu$m is 16 mJy $\pm$ 77 mJy comparable to the Herschel photometry at 70 $\mu$m of $45 ~\pm~ 16$ mJy. However at 100 microns, the IRAS flux was measured at 642 $\pm$ 280 mJy {\bf{\citep{serj2009}}}. Such high flux may be due to the large IRAS beam \footnote{The diameter of the region that would include 80\% of a point source energy is 100" at 100 $\mu$m ( Gillett, F. et al. 1985, Ch. II IRAS Catalogs and Explanatory Supplement, U.S. Government Printing Office) } which may include emission from a source 57 arcseconds away from PG 0049+171. 

{\bf{PG 0804+761}} There are two other sources in the PACS 70 $\mu$m maps within 40" of PG 0804+761's optical position. As such the high spatial resolution Herschel photometry presented here differs from previous estimates obtained from lower angular resolution data. In particular the ISO 60 $\mu$m flux measured for this source in a 45 \arcsec ~aperture is 188 $\pm$ 56 mJy, while that from MIPS at 70 $\mu$m in an 18" aperture is 110 $\pm$ 22  mJy, closer to the value of 143 $\pm$ 13 mJy from Herschel.  At 100 $\mu$m the ISO flux, also measured in a 45\arcsec ~aperture, which very likely included at least one of the two other spurious sources, is 121 $\pm$ 40 mJy, double the value of our estimate of 63 $\pm$ 13 mJy.  ISO did not detect the object at 200 $\mu$m but MIPS did at 160 $\mu$m with a value of  33 $\pm$ 6.6 mJy within a beam of  40"  \citep{mars2007}. Our PACS 160 measurement is 32 $\pm$ 8 mJy after applying the aperture correction associated with an aperture of 32". The ISO fluxes were obtained via NED from \citet{haas2003}.

{\bf{PG 1048-090}} This source is detected in all {\it{Herschel}} bands except at 160 $\mu$m. Our measurements at 70 $\mu$m agree with those from MIPS 70 as is our upper limit at 160 microns. At 160 $\mu$m there are two faint sources, neither of which are coincident with the optical source position within one PSF. A curve of growth estimate centered on the optical position of PG 1048-090 suggests a flux on the order of 7.5 mJy, or about 1.2 $\sigma$ detection. Hence we do not consider this source detected at $160~\mu $m. The SPIRE detections may be spurious sources in the beam centered at the optical position of the source. However we cannot exclude the possibility that some of the SPIRE emission may be cold dust component whose peak emission is in the 250 $\mu$m band. 

{\bf{PG 1100+772}} The 250 $\mu$m map for this source shows multiple faint sources in the central one arcminute region. It is thus possible that the 350 and 500 $\mu$m flux measurements are contaminated by emission from other sources. Given the poor angular resolution of the SPIRE 250 map and the fact that we do not detect additional sources in the higher resolution PACS data we cannot de-blend the 350 and 500 micron sources. Note however, that the centroid of the 250 $\mu$m emission is a full FWHM (18 ") away from the optical position of the source which does coincide with the PACS 70~$\mu$m emission centroid. As such we do not include the SPIRE points in our analysis and instead use our measurements as upper limits. 

{\bf{PG1114+445}} The peak of the SPIRE flux derived from the time line is only 7" away from the centroid of the 70 $\mu$m emission. However the SPIRE 250 $\mu$m map shows emission 39" away from the PACS 70 $\mu$m centroid, while the 350 $\mu$m map shows  emission 20" away from the PACS 70 $\mu$m centroid and at the same position angle as the 250 $\mu$m emission. Therefore we cannot rule out that the SPIRE photometry is contaminated by a spurious source. If the spurious source is at the same redshift as PG 1114+445, they would be 300 kpc apart. We set the values we measure at 350 and 500 $\mu$m as upper limits. 

{\bf{PG1119+120}} The 70 $\mu$m field includes four other source less than 1'  from PG 1119+120's optical position. The 160 $\mu$m flux measurement is more susceptible than those at 70 and 100 $\mu$m to contamination from those sources due to the larger PSF at 160 $\mu$m which can be seen in the larger error bar.

{\bf{PG 1211+143}} The 350 and 500 $\mu$m maps show some faint structure, and spurious sources close to the position of PG 1211+143 may have artificially increased the flux by around 20-30\% at those wavelengths based on the ratio of the faint likely spurious emission around the source and the source.   

{\bf{PG 1302-102}} The 250, 350 and 500 $\micron$ measurements are very secure as the central source in each of the maps is bright and compact. The increasing flux with wavelength, although unusual, appears to be real, and may be related to the significant contribution of non-thermal continuum. 

{\bf{PG1309+355}} Both, our PACS 70 $\mu$m estimate matches the {\it{Spitzer}} MIPS 70 $\mu$m measurement\citep{shang2011}, and our estimate at 160 $\mu$m are 59 $\pm$ 21 mJy matches those from {\it{Spitzer}} MIPS at 160 $\mu$m: 46 $\pm$ 12 mJy. We note the presence of a spurious source in the PACS 160$\mu$m map 80" away from PG1309+335.  

{\bf{PG1322+659}} At all PACS wavelengths this source is composed of two overlapping blobs (one significantly brighter than the second) the same size as our PSF, with the centroid of the combined emission within 2" of the optical position. The photometry we report here corresponds to the combined fluxes from those blobs. We also note the presence of an extended system composed of a faint ring about 17" in size, with a central unresolved blob, located at an angular distance of 35" from the optical position of PG1322+659 source in the 70~$\mu$m map. 

{\bf{PG 1341+258}} An extended (20-30") source at the edge of the image may impact the flux measured for this source because it increases the background throughout the image given the large PACS PSF \footnote{ Only 90\% of the total energy of a source is included in the inner 20-30" distance from a source. Please refer to: \url{http://herschel.esac.esa.int/twiki/pub/Public/PacsCalibrationWeb/PhotMiniScan_ReleaseNote_20101112.pdf}}. This affects the 160 $\mu$m data because of its worst angular resolution among the PACS observations. 

At 500 $\mu$m we find spurious emission from a source that is 1' away from the PACS 70 $\mu$m emission centroid. We use the measured emission to estimate upper limits at 500 $\mu$m. 

{\bf{PG 1354+213}} The SPIRE source we detect in the vicinity of our source ($\sim$ 13" away from the 70$\mu$m emission centroid) does not appear to be associated with PG 1354+213 both because of their positions and also because they are a factor of a few brighter than the emission at 160$\mu$m. We use our time-line data measurement to determine an upper limit for the SPIRE emission associated with the QSO. 

{\bf{PG 1612+261}} The 500 $\mu$m measurement from the time-line is higher than what would be expected from the modified black-body that fits the PACS measurements as well as the 250 and 350 $\mu$m data points (see section 3.4). Since the SPIRE 500 $\mu$m map does not show a conclusive detection we use the time-line estimate as an upper-limit, but cannot rule out that this source has a cold dust emission component in excess of what can be determined from a modified black-body fit to the 70-350 $\mu$m measurements. 

{\bf{PG 1704+608}} The 250 and 350 $\mu$m emission centroids are within a PSF away from 70 $\mu$m emission centroid, however the 500 $\mu$m emission center is $\sim$ 36" away, i.e. about 1 FWHM. For reference that source at 500 $\mu$m is 33 $\pm$ 9 mJy and we use this measurement to derive an upper limit at 500 $\mu$m. 

{\bf{PG 2209+184}} The 500 $\mu$m point from the time-line fits is higher than what would be expected from the modified black-body that fits the PACS measurements as well as the 250 and 350 $\mu$m data points (see section 3.4).  In the 250 and 350 $\mu$m maps we see 3 sources within 80" of the position of the source, note that is is more than the PSF width at 500 $\mu$m of $\sim$37".  Therefore we cannot conclusively determine whether the high value of the estimated flux at 500 $\mu$m is due to contamination from the sources around it or an excess of cold dust in the host galaxy.  %We thus use the time-line estimate 

{\bf{PG 2304+042}} The SPIRE maps are dominated by several sources as well as faint extended emission around the central area, perhaps associated with a foreground object. The morphologies of the SPIRE detections do not match the morphology of the faint 70 $\mu$m emission. The timeline fitting at the optical position of PG2304+042 results in detections at 250 and 350 $\mu$m which we present in Table 3. However the SPIRE emission we measure may be spurious and not physically associated with the source, but we do present them as detections as hey are located at the correct optical position and are above 3$\sigma$.

\section{A2: Synchrotron versus Thermal Dust Emission}
We investigate the radio properties of the PG QSOs to estimate the amount of synchrotron radiation which contributes to the observed FIR emission in radio loud PG QSOs.

We compile observations of PG QSOs at 1.4 GHz from the Faint Images of the Radio Sky at Twenty-cm \citep[FIRST]{bwh1995}, the NRAO Very Large Array Survey \citep[NVSS]{condon1998}, and 5 GHz VLA data  from \citep{kel1989}. The FIRST survey covered 10,000 square degrees with a 1.0 mJy source detection threshold and a spatial resolution of 5".  NVSS covers the entire sky north of -40 deg declination to a completeness limit of 2.5 mJy and a spatial resolution of 45". The observations of \citet{kel1989} used to estimate total fluxes at 5 GHz reached a sensitivity of 0.065 mJy and a spatial resolution of 18". Note that \citet{kel1989} observed a subset of the PG QSOs at a higher resolution of 0.5" to determine the strength of unresolved components. Here we only use the 18" resolution data. 

Two definitions are used to demarcate radio-quiet and radio loud QSOs. One criterion is based on the ratio of optical to radio flux $R$. As defined and measured by \citep{kel1989}, $R$ is the ratio of the total flux density at 6 cm to the optical flux density at an effective wavelength of 4400 \AA.  Using this definition and the $R$ estimates from \citep{kel1989} 16 PG QSOs from our sample are radio loud. 
 \citet{pea1986} point out that $R$ can be used as a discriminating parameter only if the radio and optical luminosities are linearly correlated, which is not the case \citep{sto1992}. Note however, that on the basis of a radio-selected sample of 636 QSOs \citet{white2000} have argued that the distribution of radio characteristics is not bimodal. 
 
The second definition \citep{greg1996} divides sources at the 1.4 GHz luminosity of $3~\times~ 10^{25}$ W/Hz. \citet{ive2002} found that the two definitions are consistent for optically selected QSOs. 
We compute the rest frame luminosities at 1.4 GHz assuming a spectral index of -0.5 \citep{Stern2000} and find that $\sim13\%$ are brighter at 1.4 GHz than $10^{25}$ W/Hz and would qualify as radio loud AGN under the definition of \citep{greg1996}.  

We estimate the possible contributions of synchrotron emission at 250 $\mu$m using the example of \citet{kal2014} using steeper spectral index of  -0.7.   We find that only in 13  sources the contribution from synchrotron could exceed 10 mJy - the typical error on the measurements at 250 $\mu$m. However we note that only 7 of the PG QSOs have actually been detected at wavelengths longer than 250 $\mu$m.  

To assess if and how our investigation of FIR properties of  this sample of Type 1 QSOs is biased by the inclusion of radio loud QSOs we compare their FIR to their radio properties. Figure \ref{LradVdust} shows S$_{250\mu m}/\rm{S}_ {24\mu m}$,  S$_{350\mu m}/\rm{S}_ {250\mu m} $ and L$_{11.3 \mu m \rm{PAH}}/\rm{L}_{\rm{FIR}}$ as a function of the 1.4 GHz radio luminosity and the radio loudness factor $R$. We do not find any obvious trends between the IR slope and the radio luminosity, or $R$ in agreement with the work of \citep{kal2014} and so our results in the FIR do not appear biased by the radio loud sources in the sample.

\clearpage
%PUT ALL THE FIGURES AT THE END HERE
\begin{figure*}[!h]
$\begin{array}{cc}
\includegraphics[angle=0,width=0.49\linewidth]{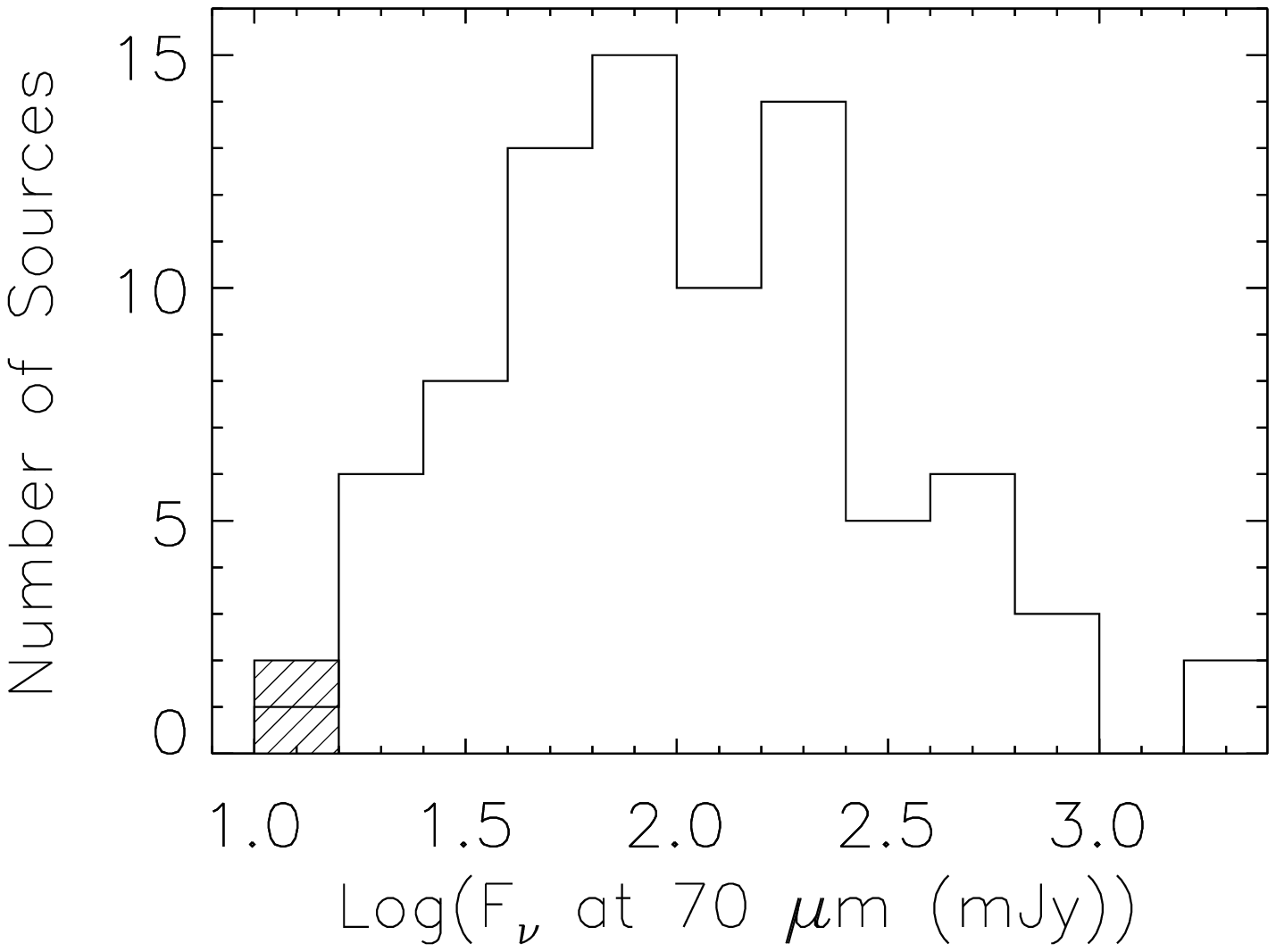}
\includegraphics[angle=0,width=0.49\linewidth]{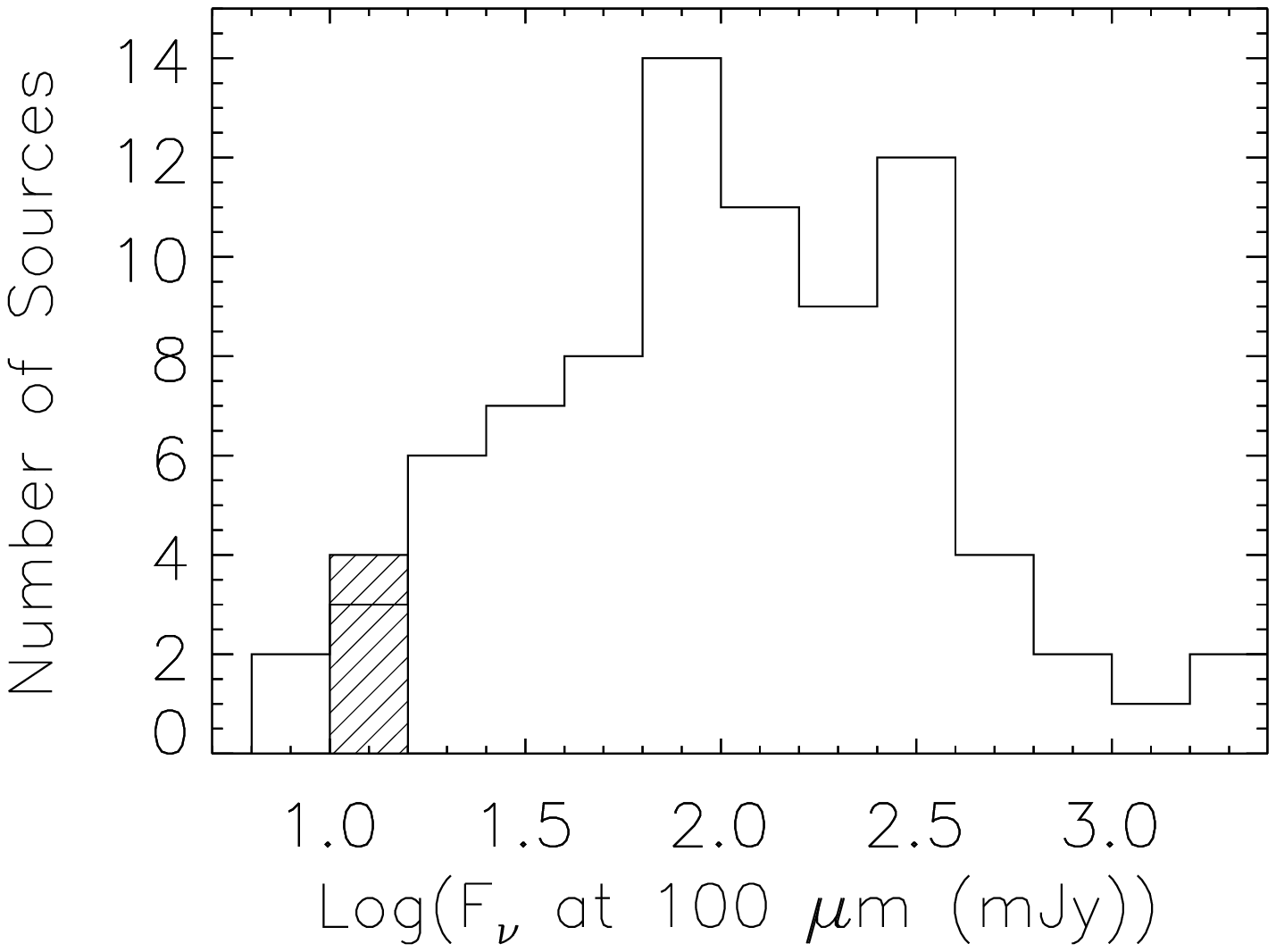}\\
\includegraphics[angle=0,width=0.49\linewidth]{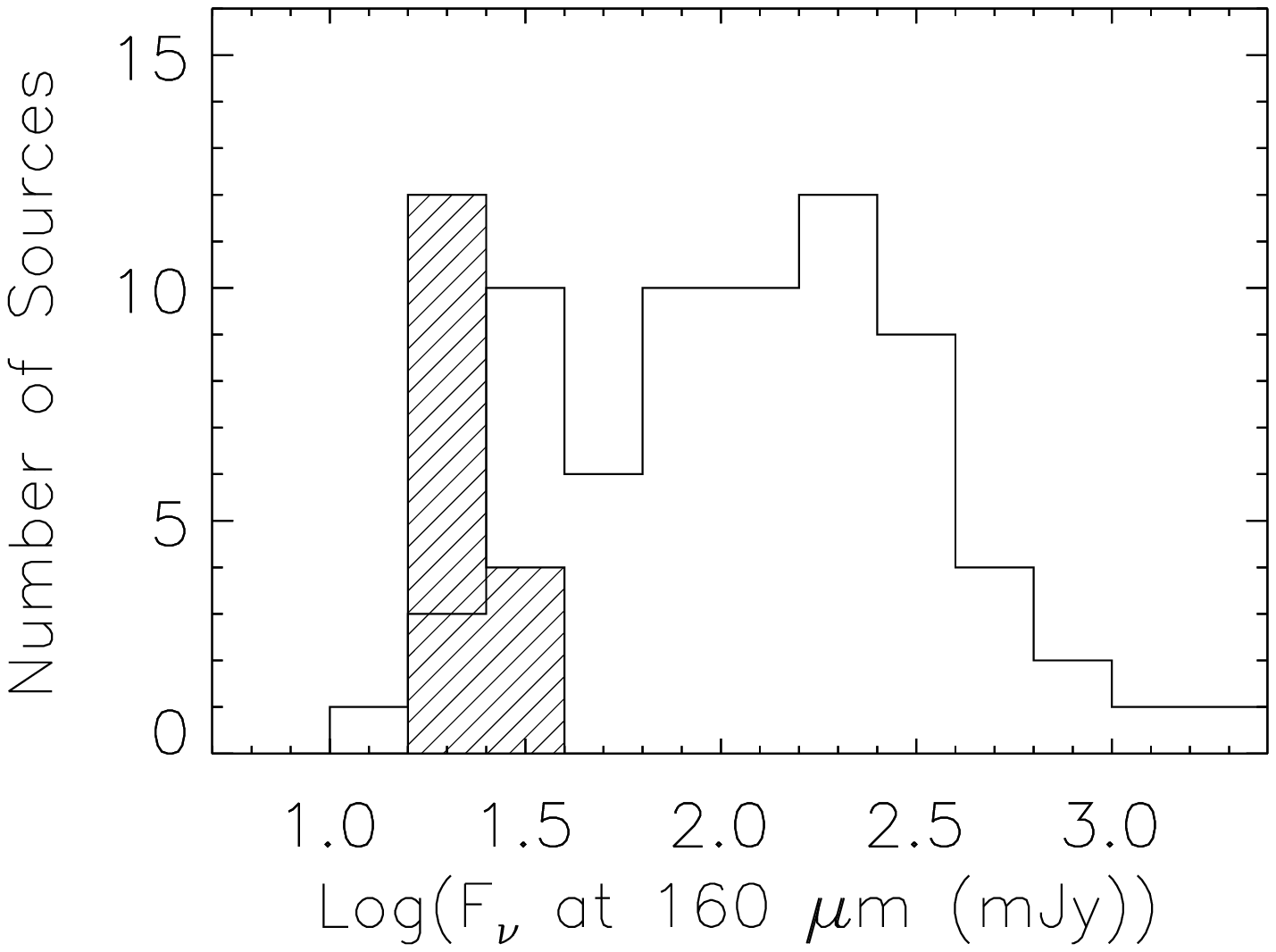}
\includegraphics[angle=0,width=0.49\linewidth]{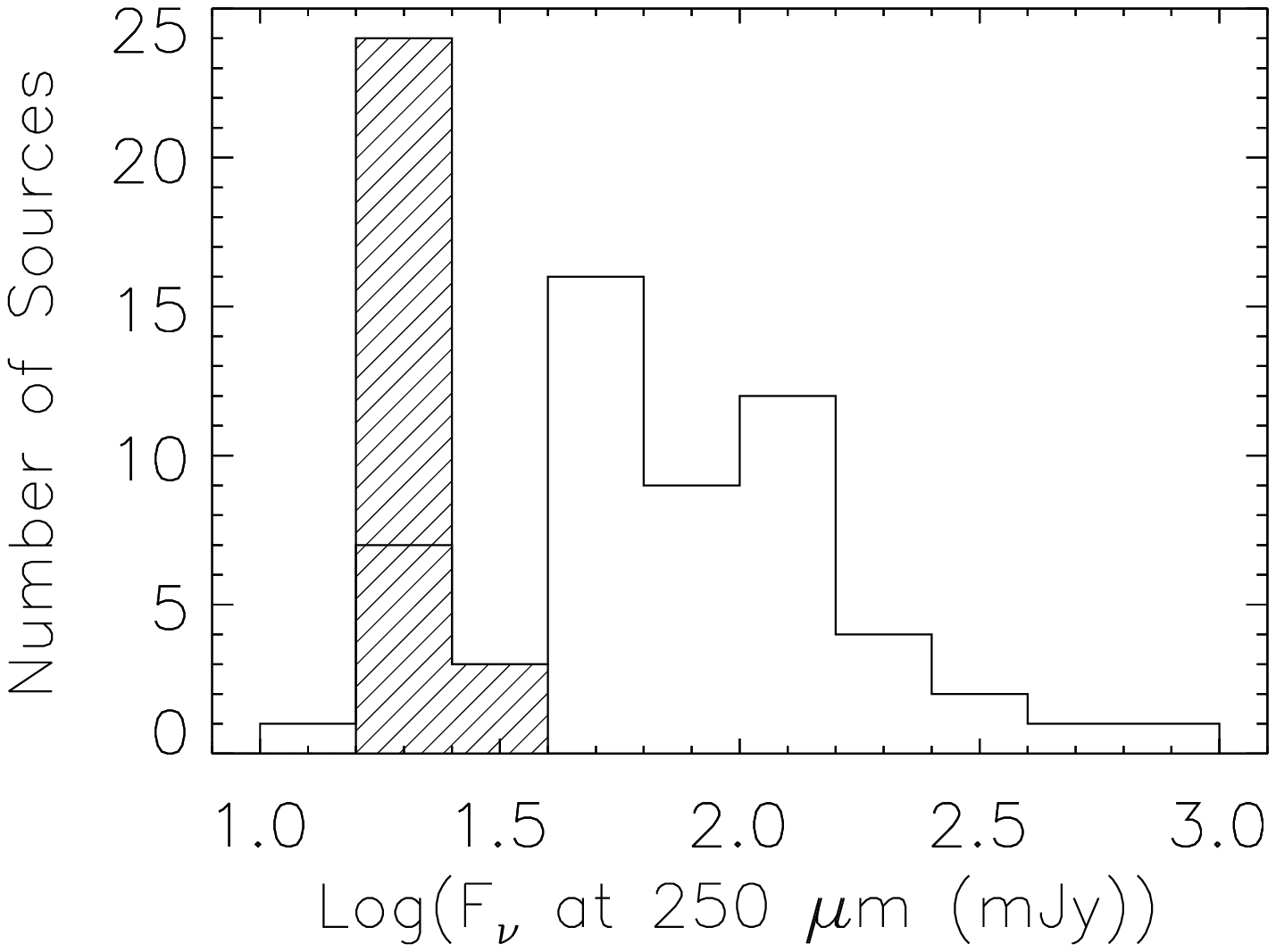}\\
\includegraphics[angle=0,width=0.49\linewidth]{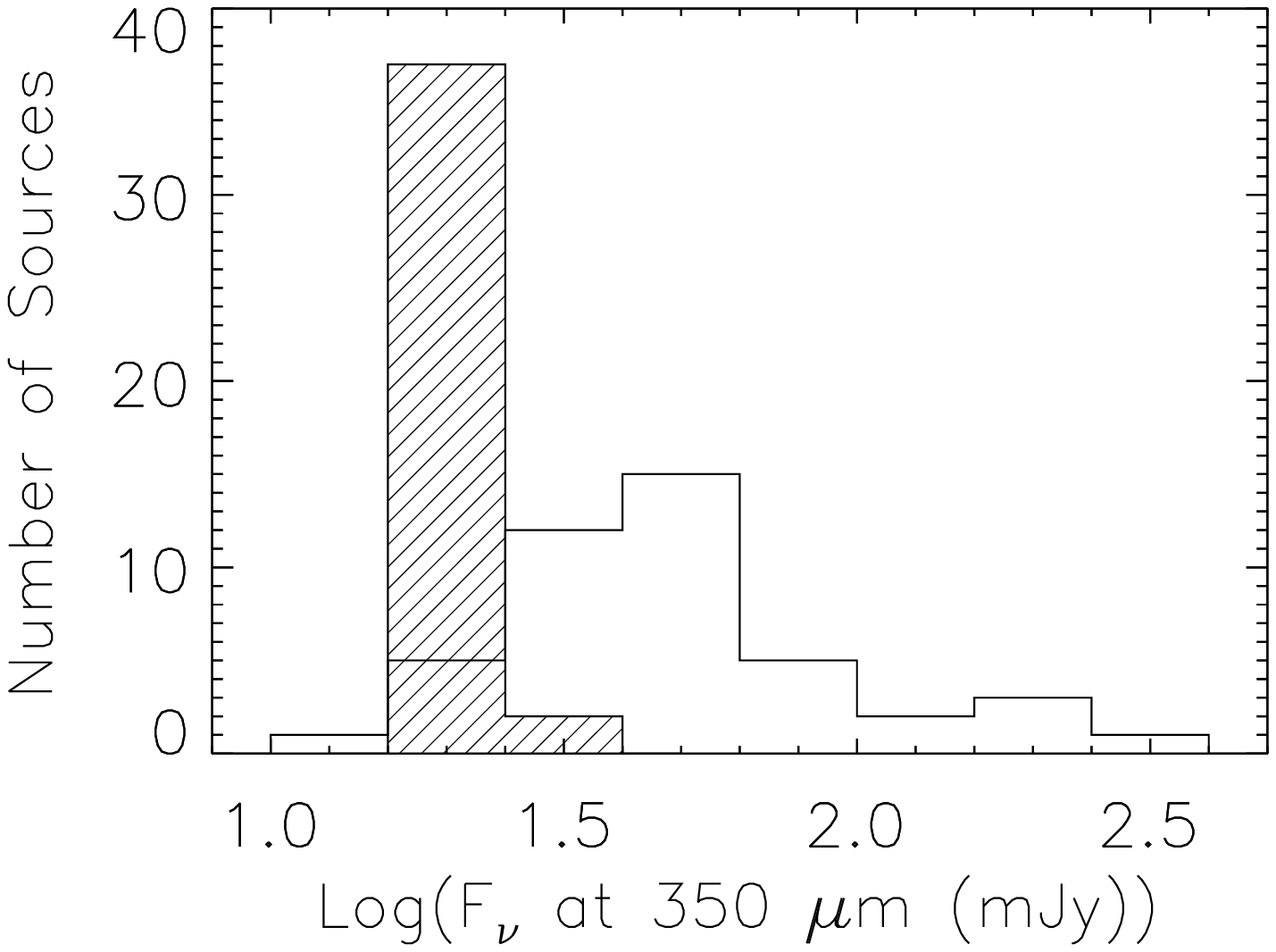}
\includegraphics[angle=0,width=0.49\linewidth]{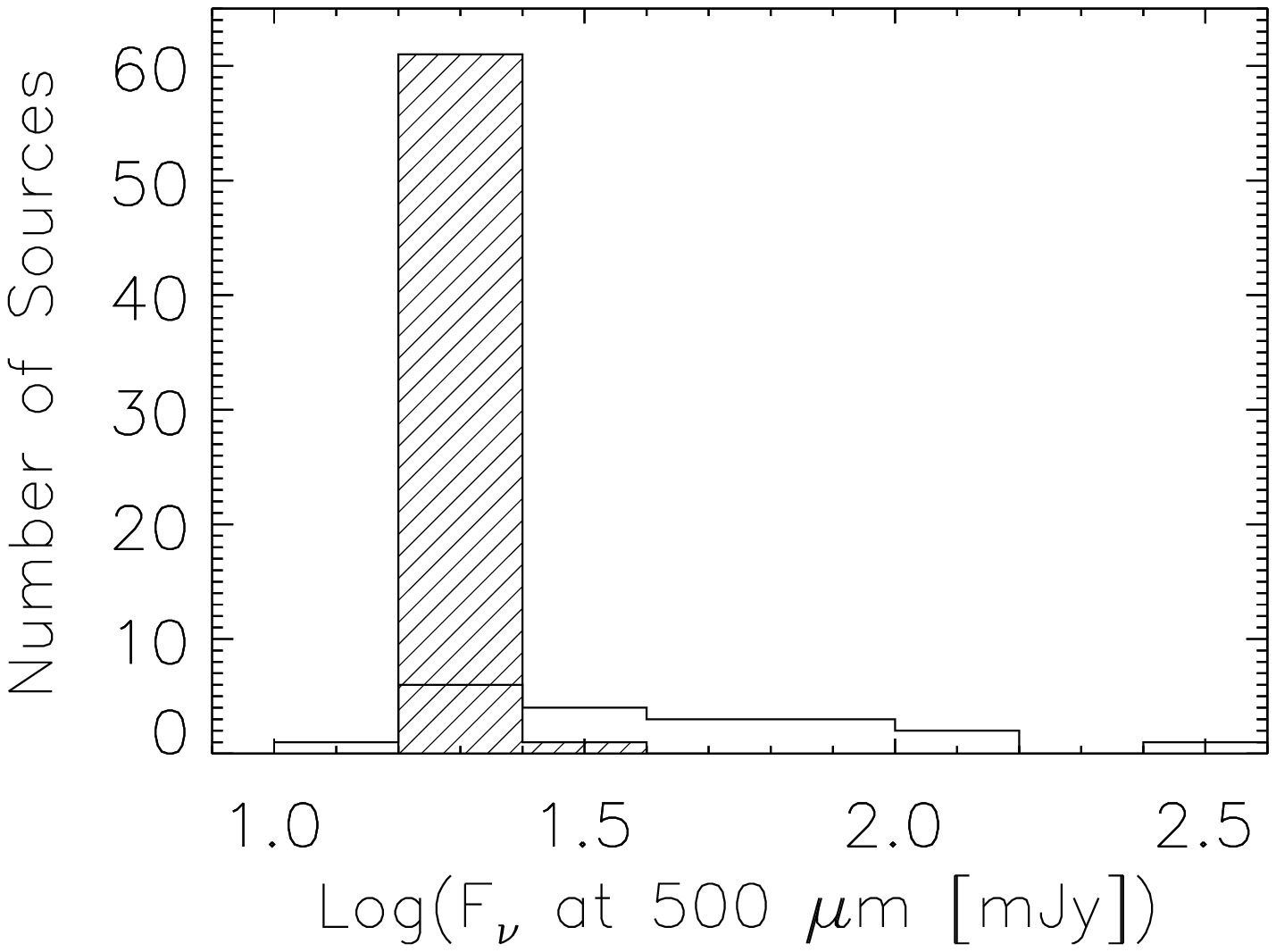}\\
\end{array}$
\caption{{\small{Histograms observed Herschel PACS and SPIRE fluxes at 70, 100, 160, 250, 350, and 500 $\mu$m. The upper limits are shown as dashed histograms. }} \label{FLnLUMS}}
\end{figure*}

\begin{figure*}[!h]
$\begin{array}{cc}
\includegraphics[angle=90,width=0.49\linewidth]{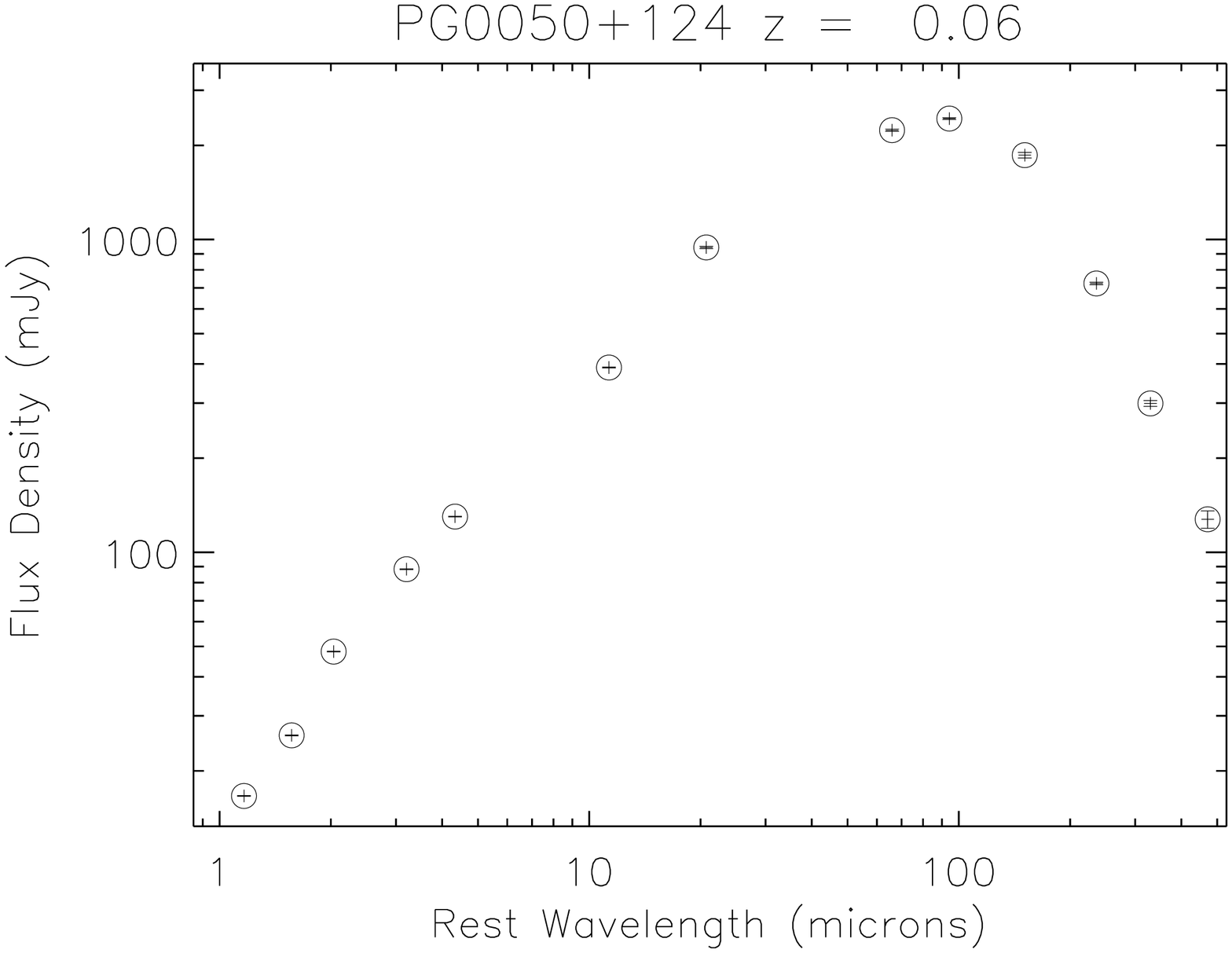}
\includegraphics[angle=90,width=0.49\linewidth]{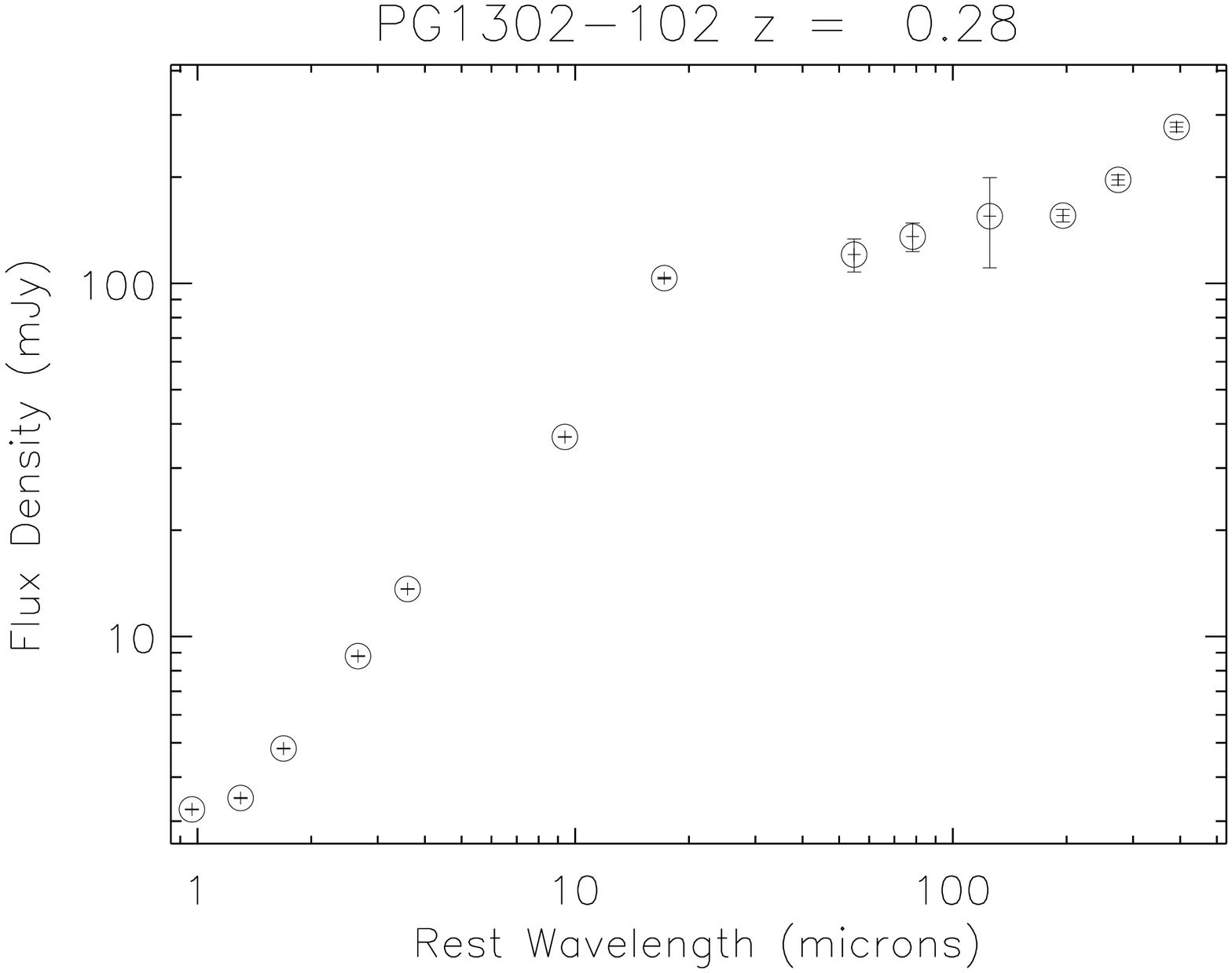}\\
\end{array}$
\caption{NIR to FIR SEDs of PG 0050+124 and radio loud source PG 1302-102.  All the SEDs are available in the electronic version of this paper. \label{PSEDs}}
\end{figure*}

\begin{figure*}[!h]
 \includegraphics[angle=90,width=0.6\linewidth]{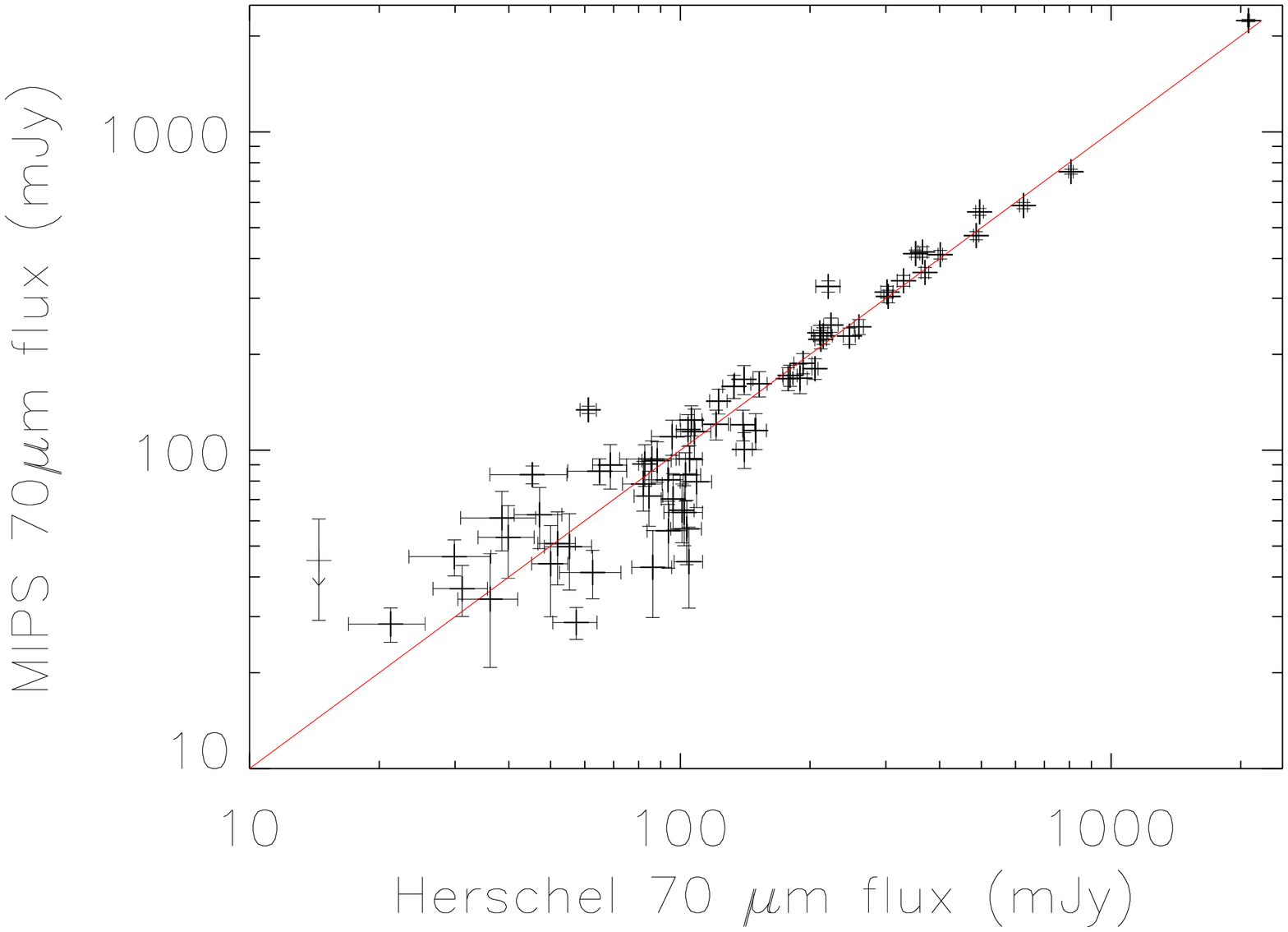}\\
 \includegraphics[angle=90,width=0.6\linewidth]{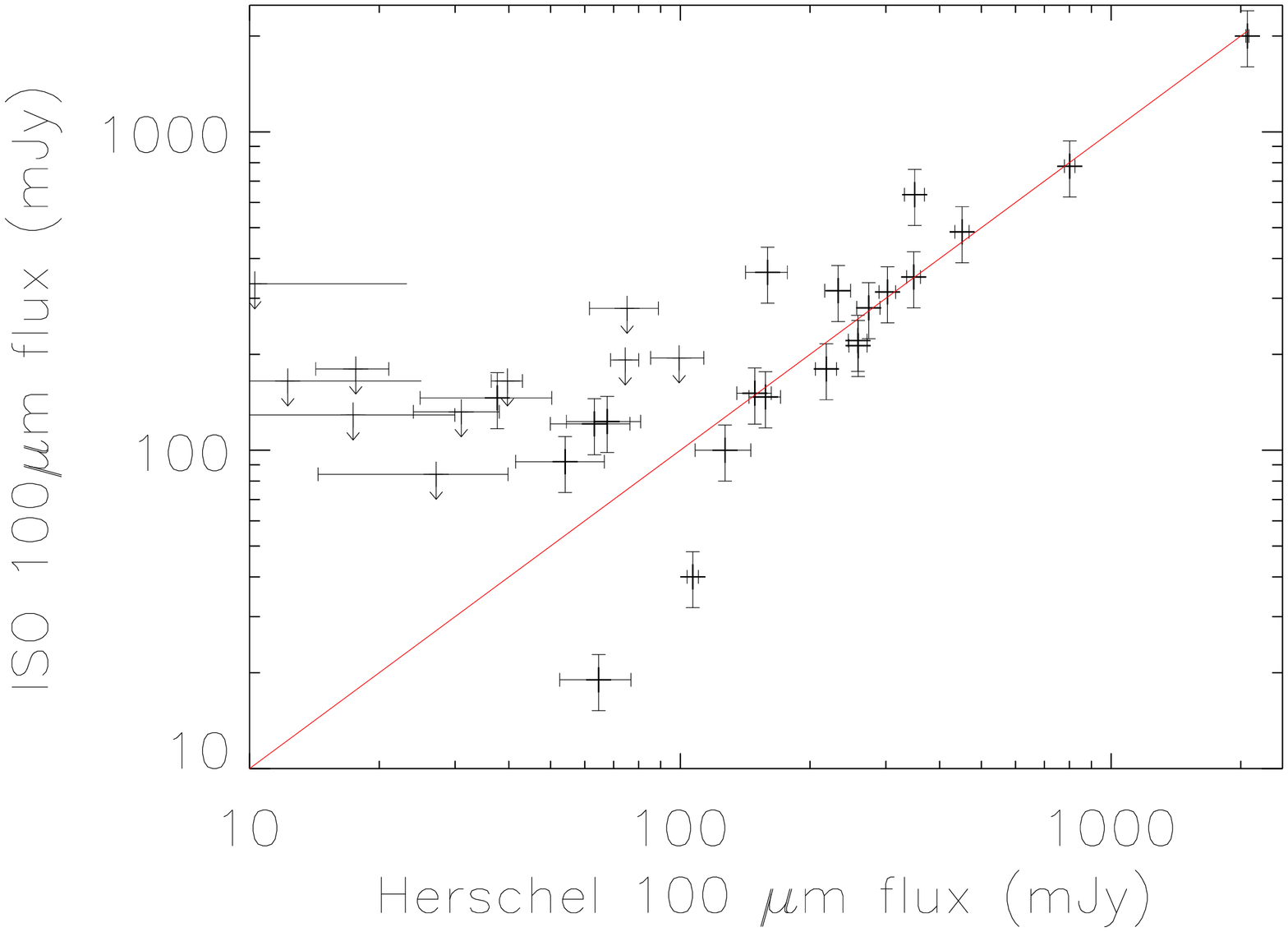}\\
  \includegraphics[angle=90,width=0.6\linewidth]{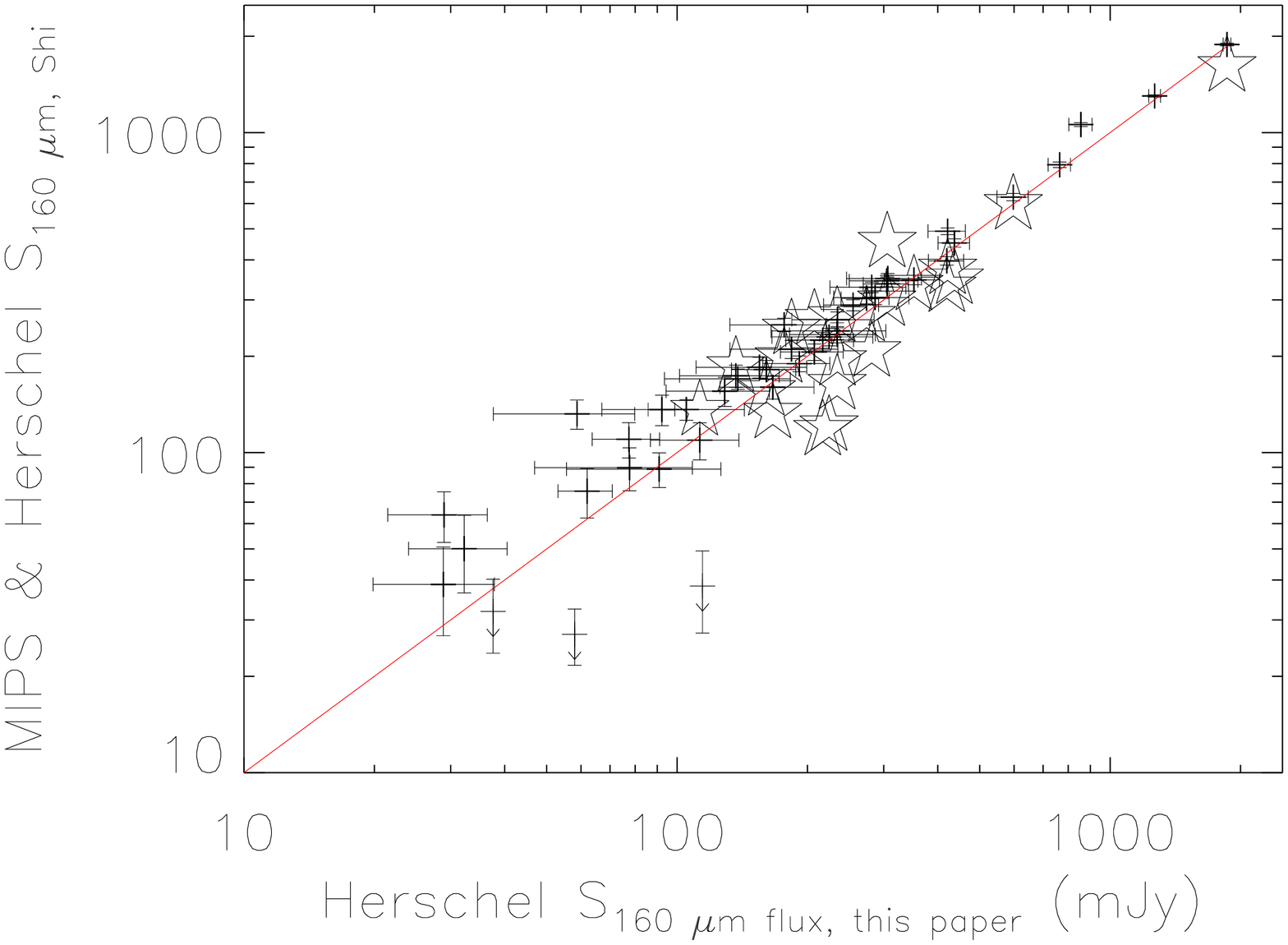}\\
 \caption{Herschel measurements at 70, 100 and 160 $\mu$m presented in this paper versus MIPS 70$\mu$m measurements from \citet{shi2014}, ISO 100 $\mu$m data from ISO \citet{haas2003} and MIPS (stars) and Herschel (crosses) measurements from \citet{shi2014} \label{iso100}}
\end{figure*}

\begin{figure*}[!h]
\includegraphics[angle=90,width=0.49\linewidth]{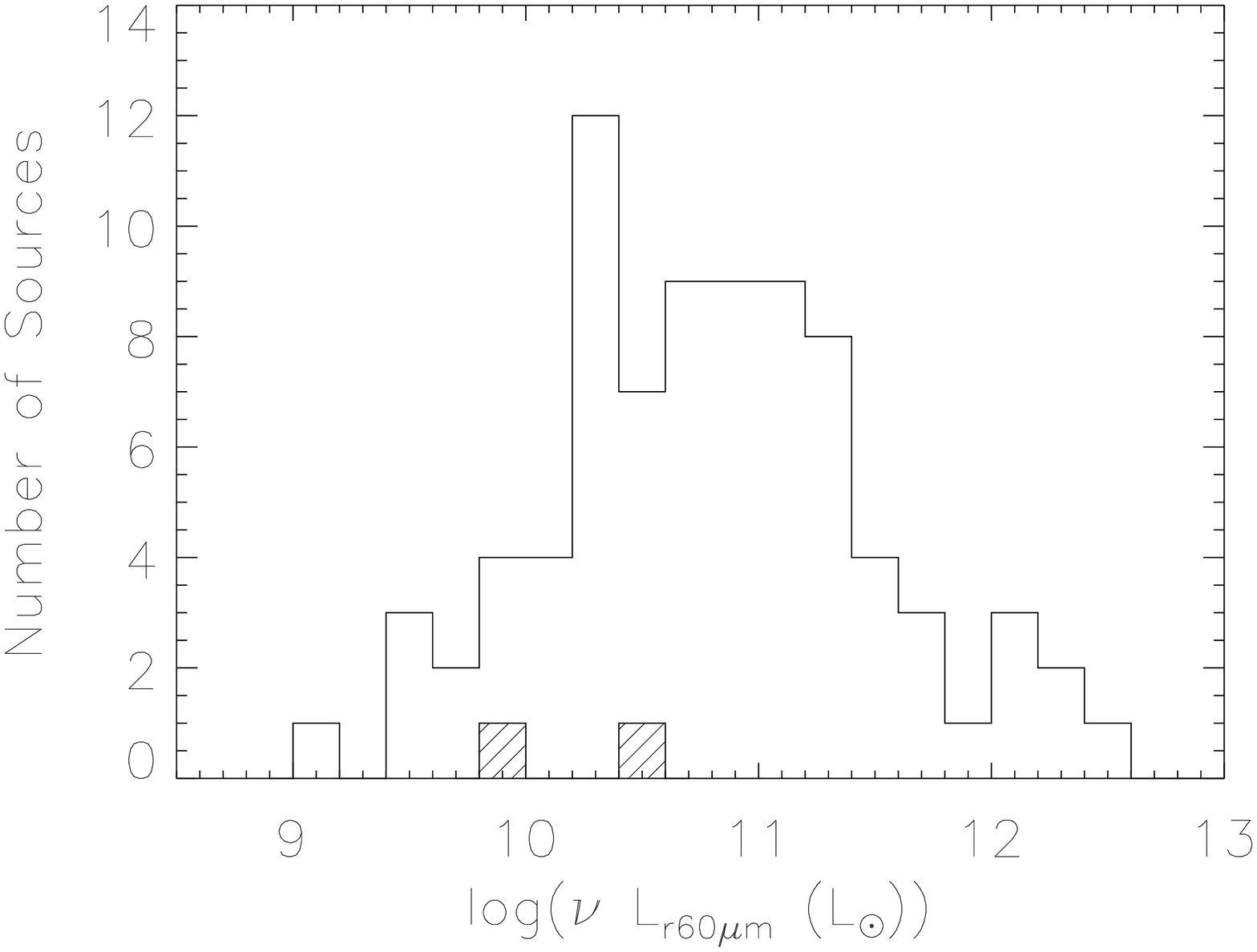}
\includegraphics[angle=90,width=0.49\linewidth]{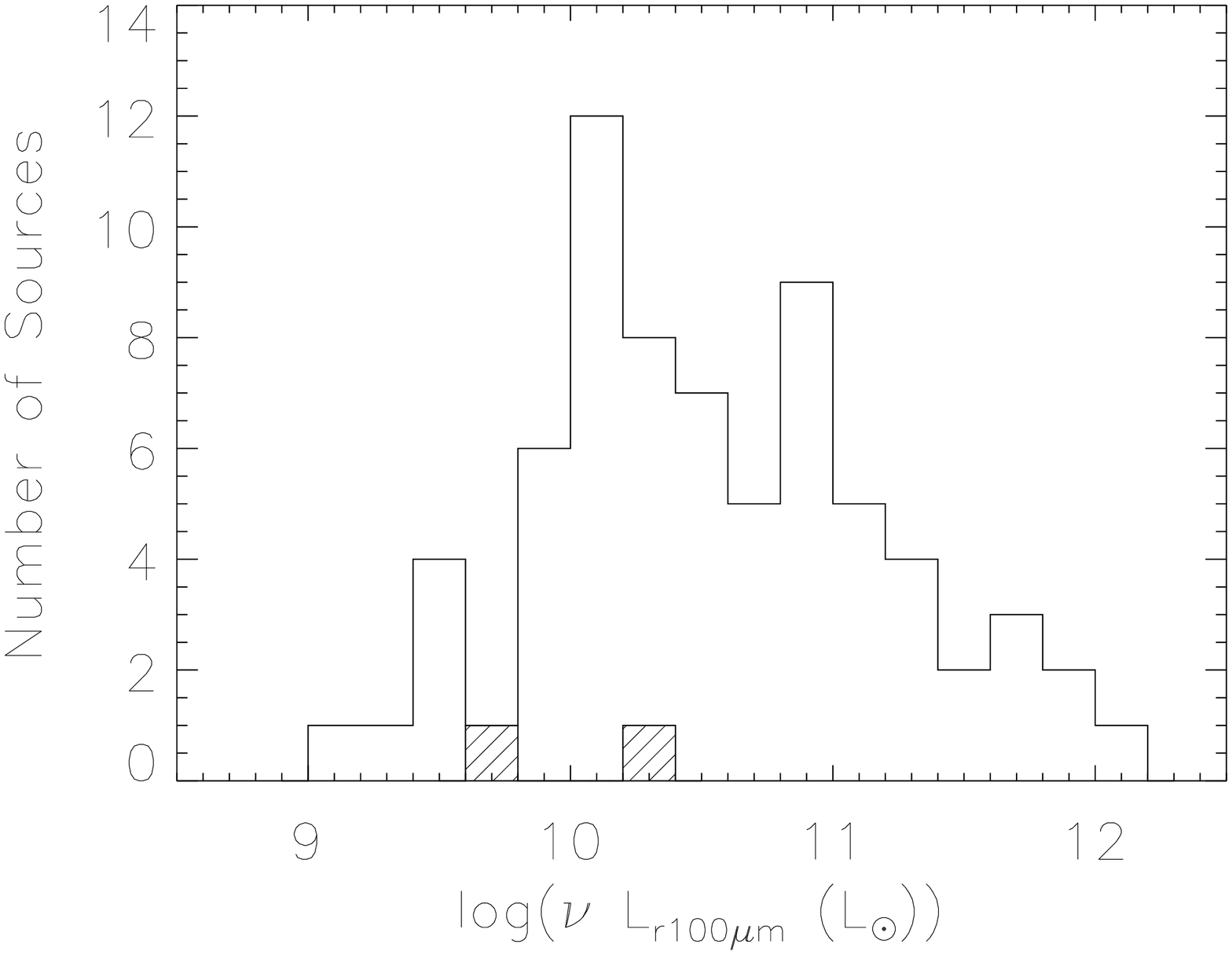}\\
\includegraphics[angle=90,width=0.49\linewidth]{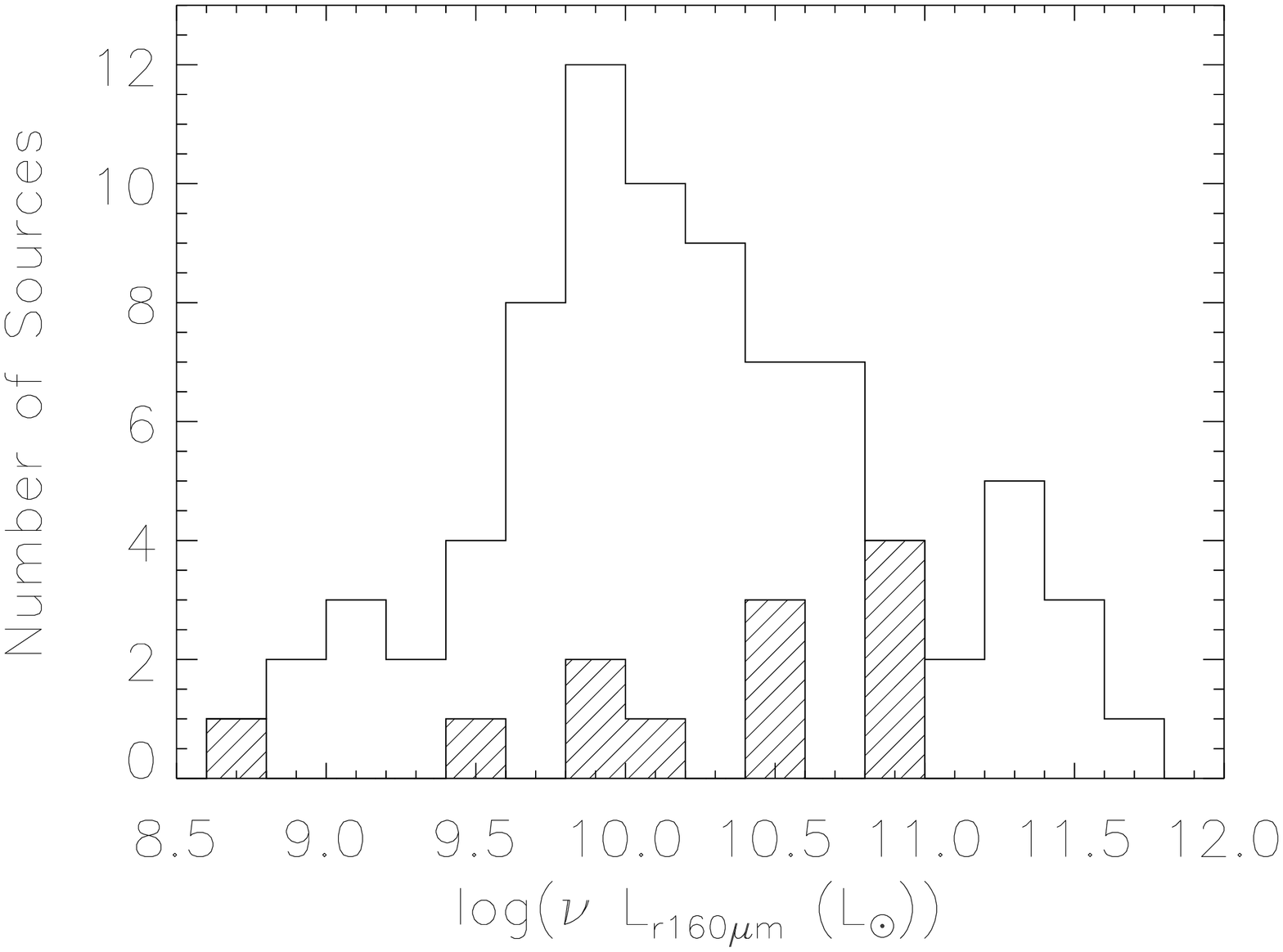}
\includegraphics[angle=90,width=0.49\linewidth]{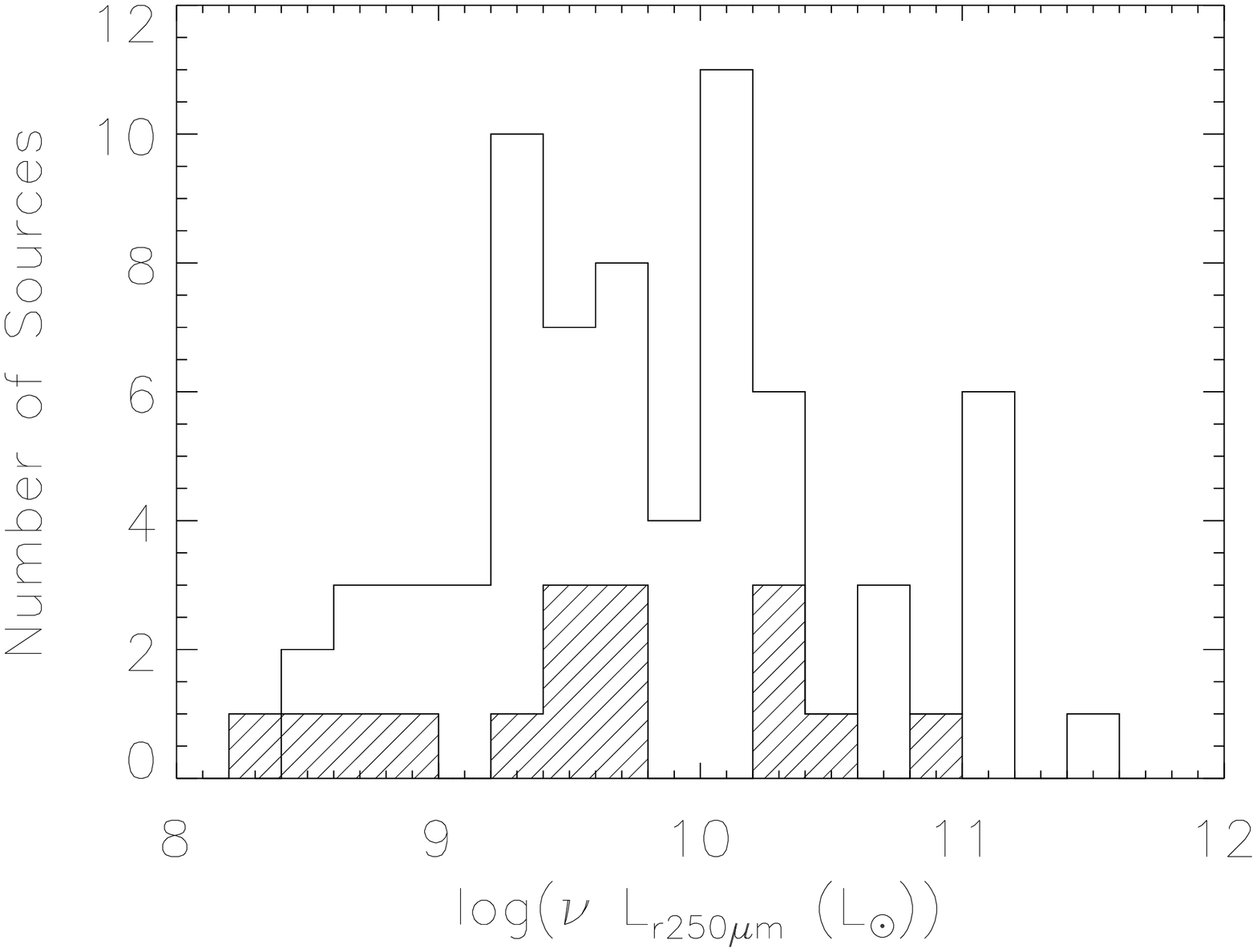}\\
\includegraphics[angle=90,width=0.49\linewidth]{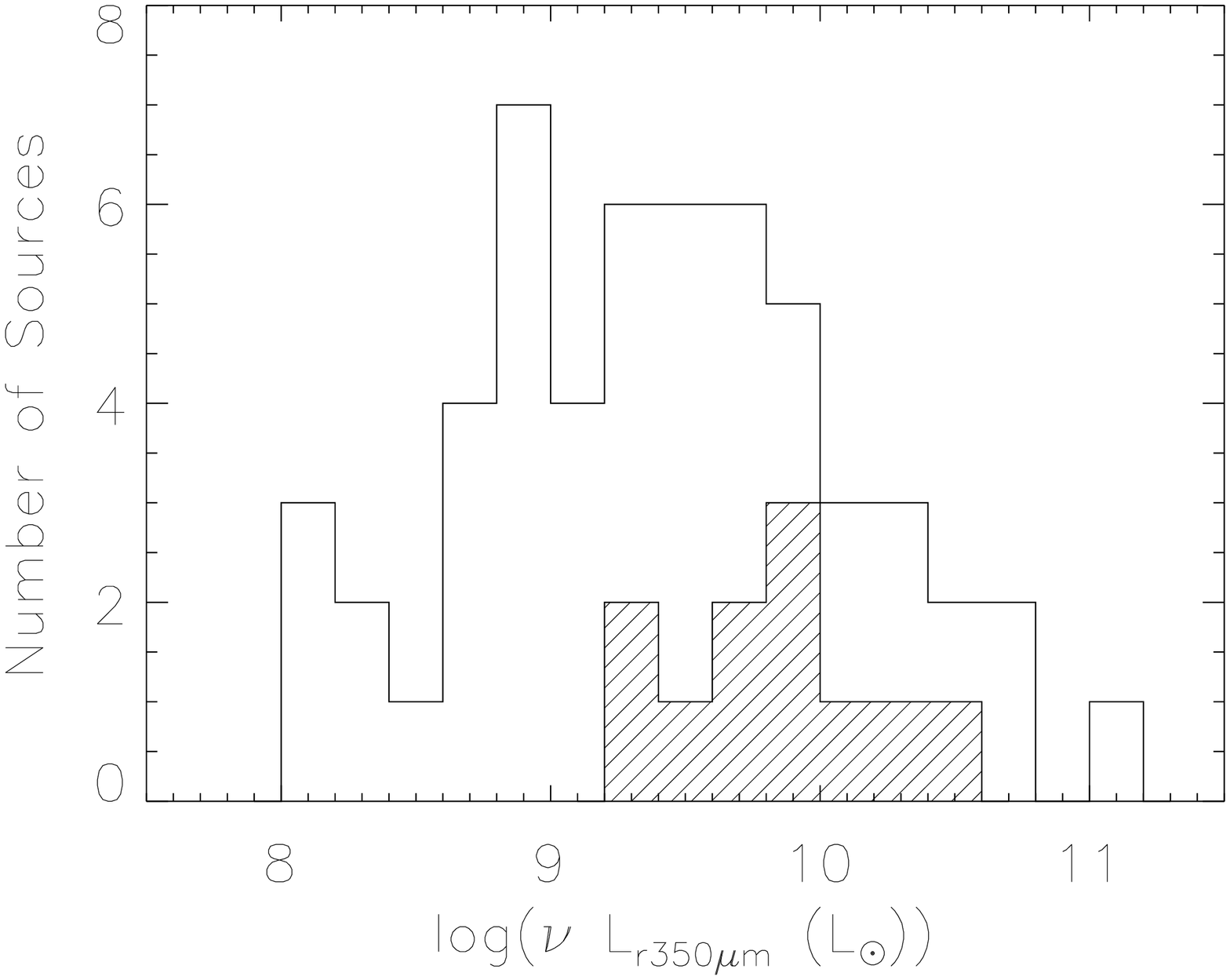}
\includegraphics[angle=90,width=0.49\linewidth]{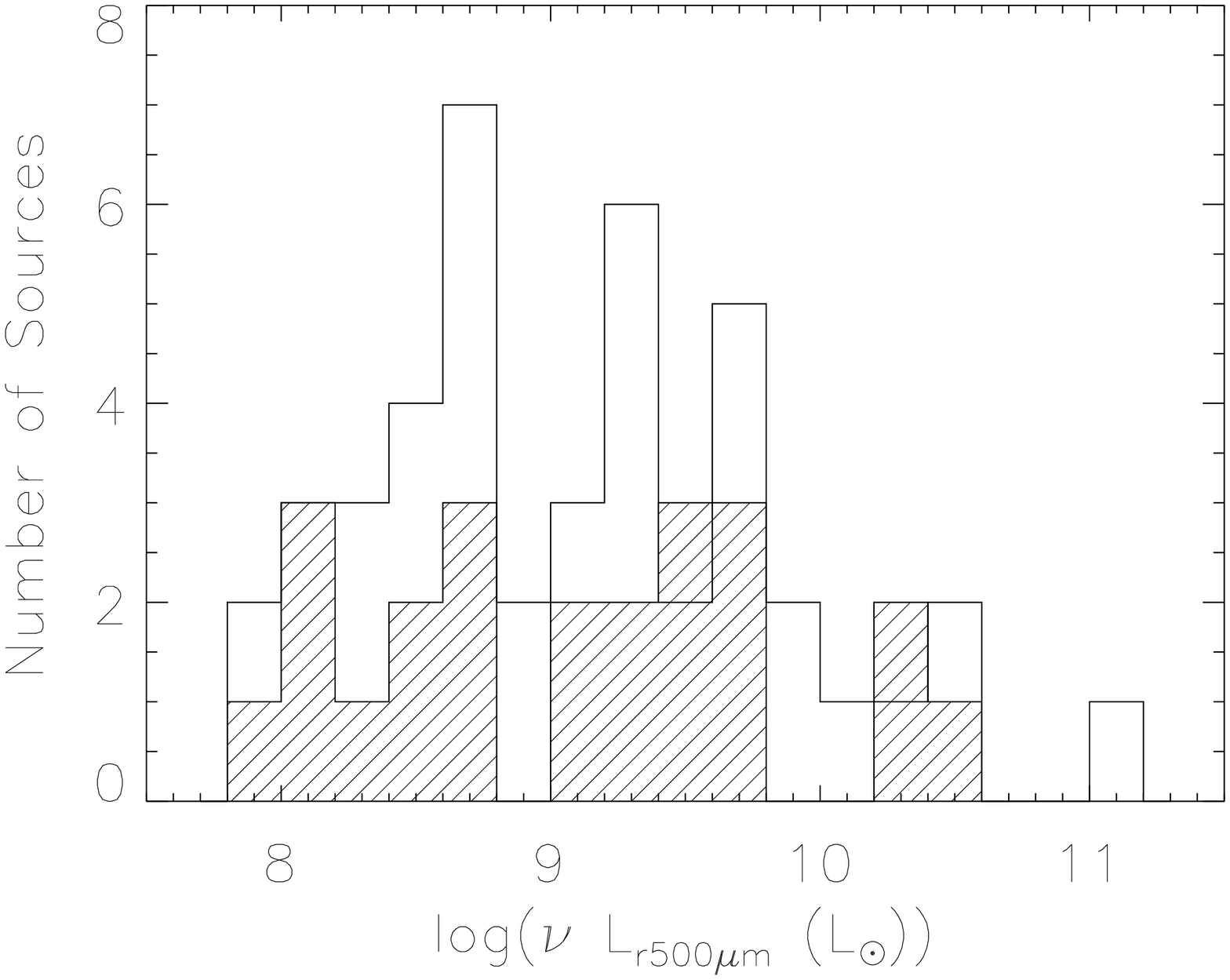}\\
\caption{Histograms of rest-frame luminosities at 60-500 $\mu$m obtained from {\it{Herschel}} photometric measurements. The upper limits are shown as dashed histograms.  \label{Rlums}}
\end{figure*}

\begin{figure*}[!h]
\includegraphics[angle=90,width=0.9\linewidth]{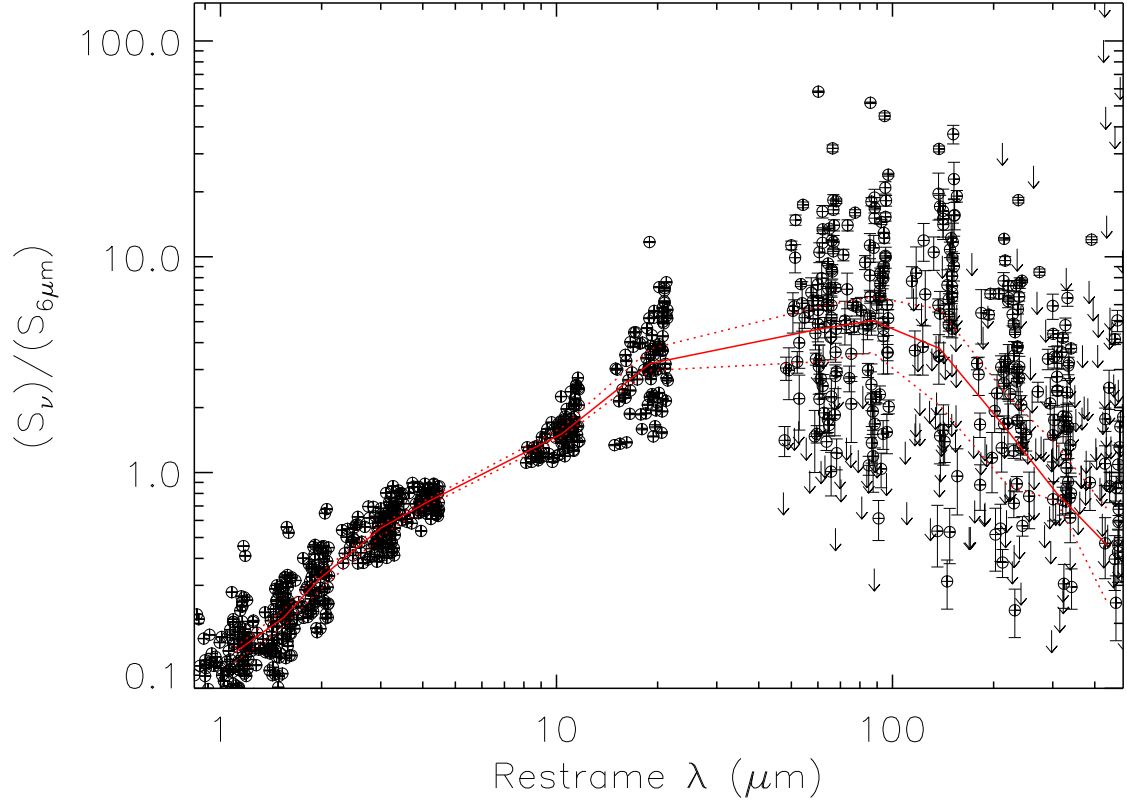}
\caption{The combined photometry of all the 85 sources in our sample normalized by the estimated luminosity at 6 $\mu$m. Top: solid line gives the median template computed in each observed photometric band using the Kaplan-Meier estimator, the dotted lines show the 95\% confidence interval. \label{Temps}}
\end{figure*}

%\begin{figure*}[!h]
%\includegraphics[angle=90,width=0.48\linewidth]{TemplateNetcomp.eps}
%\includegraphics[angle=90,width=0.48\linewidth]{TemplateMULcomp.eps}\\
%\caption{The combined photometry of all the 85 sources in our sample normalized by the estimated luminosity at 6 $\mu$m in units of $\lambda \rm{S}_{\lambda}$ (top) and S$_{\nu}$ (bottom). The solid lines give the median template computed in each observed photometric band using the Kaplan-Meier estimator, the dotted lines show the 95\% confidence interval in both. We compare the templates based on the Herschel photometry presented in this paper with those derived from Spitzer and ISO/IRAS data by \citet{net2007} and \citet{mul2011} in the top an bottom panels respectively.   \label{TempsC}}
%\end{figure*}

\begin{figure*}[!h]
\includegraphics[angle=90,width=0.9\linewidth]{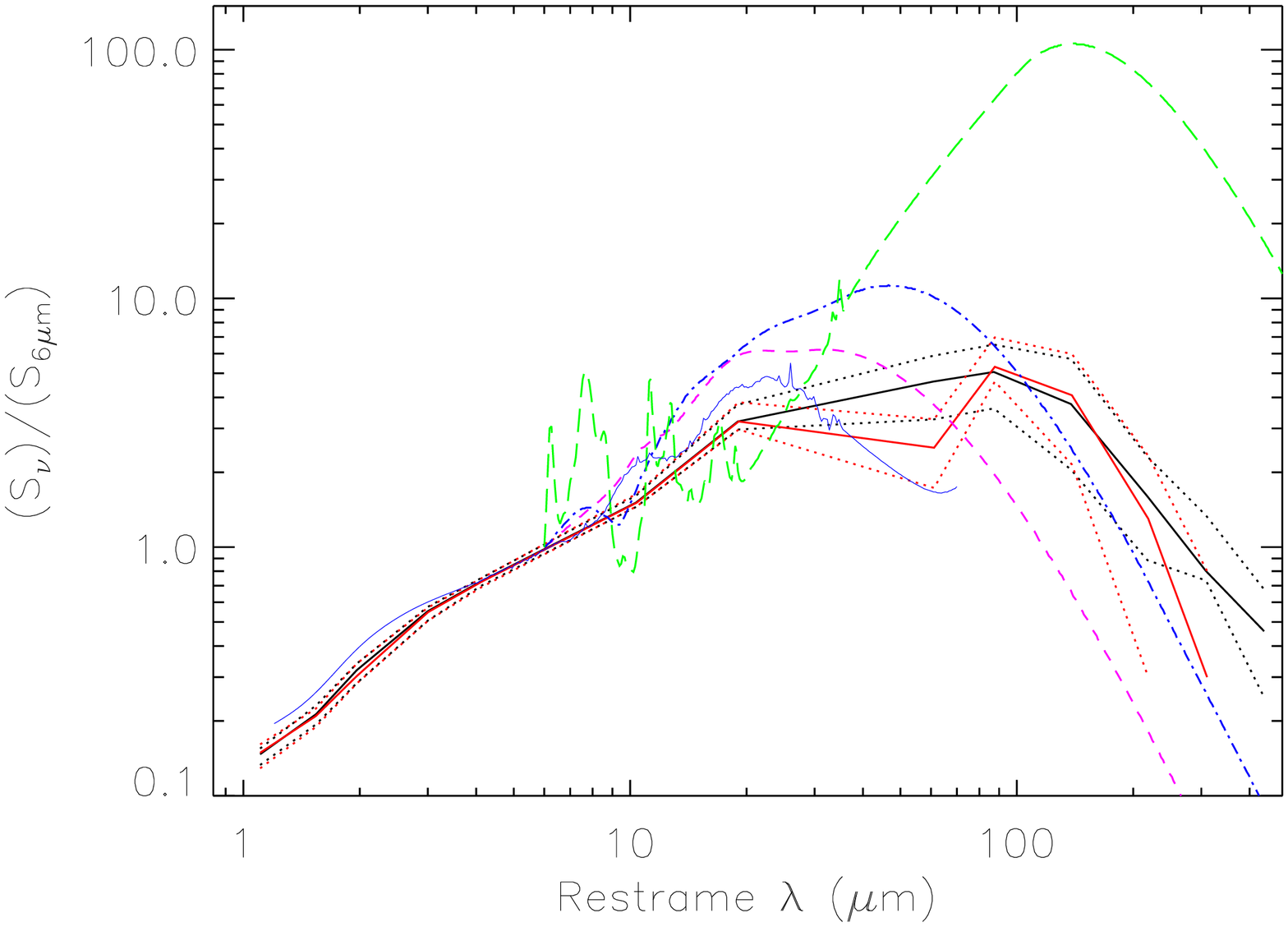}
\caption{Comparison of median SEDs derived from (1) the entire PG QSOs sample presented in this paper (black lines), (2) only the radio quiet sources (red lines), (3) starburst template from \citet{mul2011}, (4) average observed SED of 28 PG QSOs observed with IRS from \citet{net2007} (magenta solid) (5) average intrinsic AGN SED of 8 'FIR-weak' PG QSOs, after subtraction of a starburst component from \citet{net2007} (blue solid), (6) average high-luminosity: log(L$_{2-10 keV} \geq 42.9$), nearby AGN template (dashed magenta) from \citet{mul2011}, and (7) average low-luminosity: log(L$_{2-10 keV} \geq 42.9$), nearby AGN template (dot-dash blue line) from \citet{mul2011}. See Fig. \ref{Temps} for photometry of individual PG QSOs normalized at 6 $\mu$m.\label{TempsC}}
\end{figure*}

\begin{figure*}[!h]
$\begin{array}{cc}
\includegraphics[angle=0,width=0.48\linewidth]{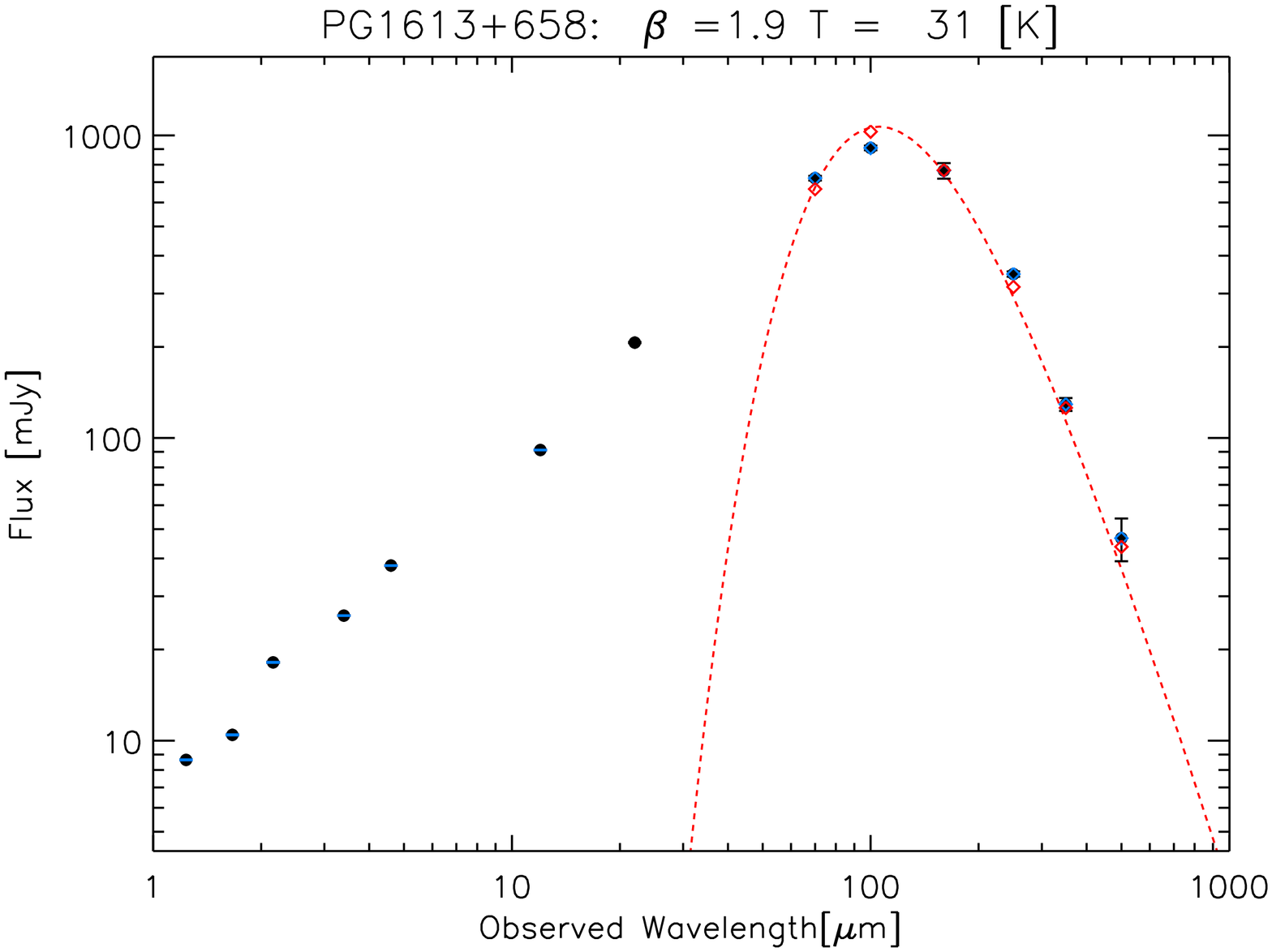}
\includegraphics[angle=0,width=0.48\linewidth]{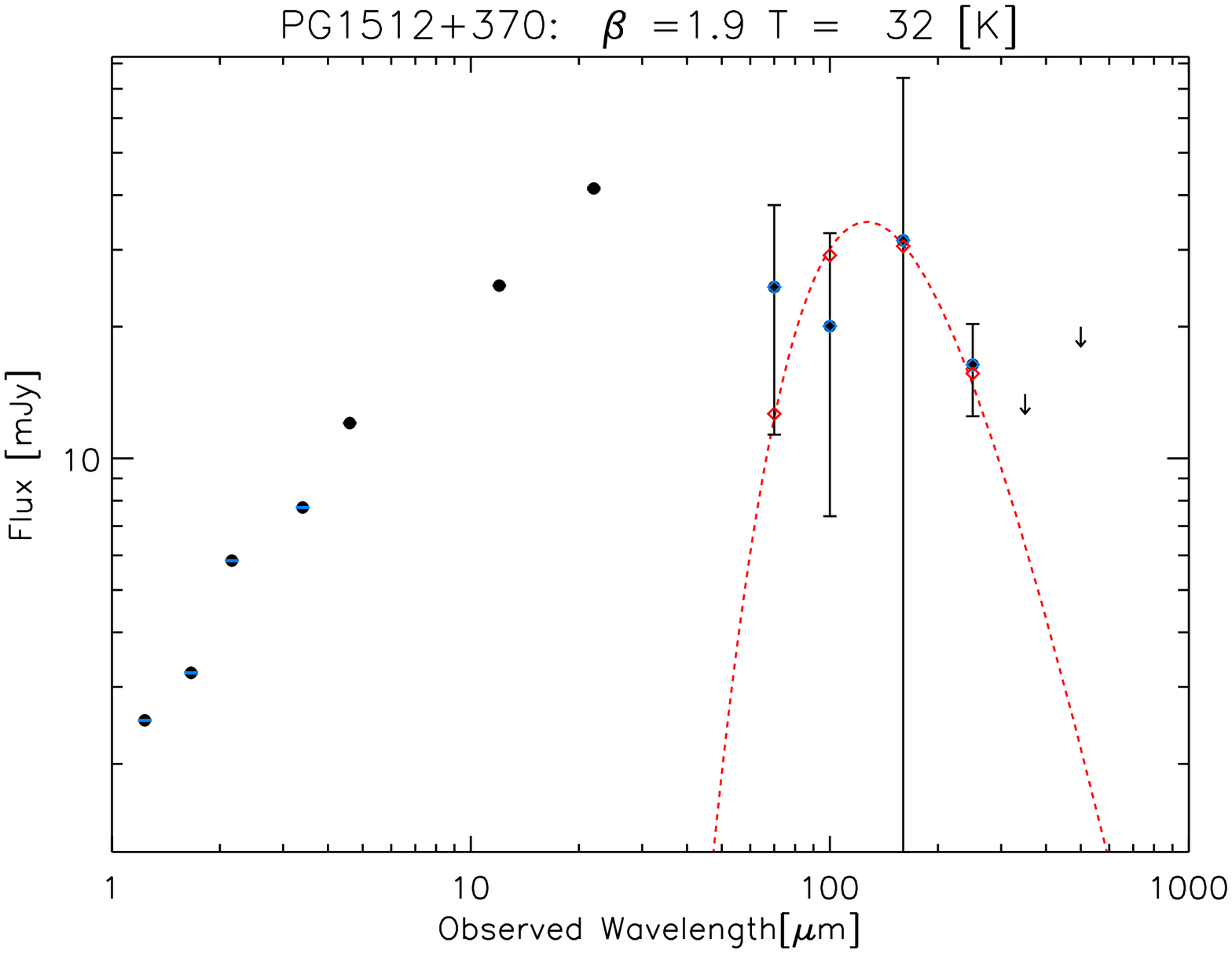}\\
\end{array}$
\caption{\label{exafit}Examples of good and poor fits. Less than 10\% of the sources have such poor fits, and those are poor because of low signal to noise ratios on individual measurements.}
\end{figure*}

\begin{figure*}[!h]
\includegraphics[angle=90,width=0.85\linewidth]{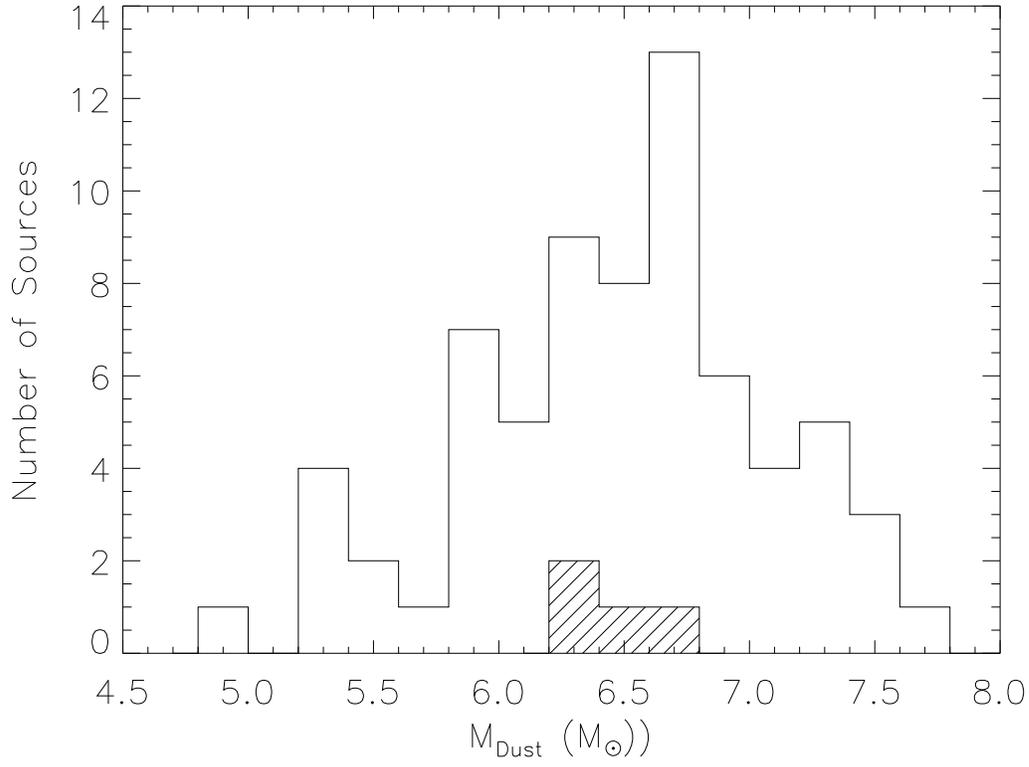}\\
\includegraphics[angle=90,width=0.85\linewidth]{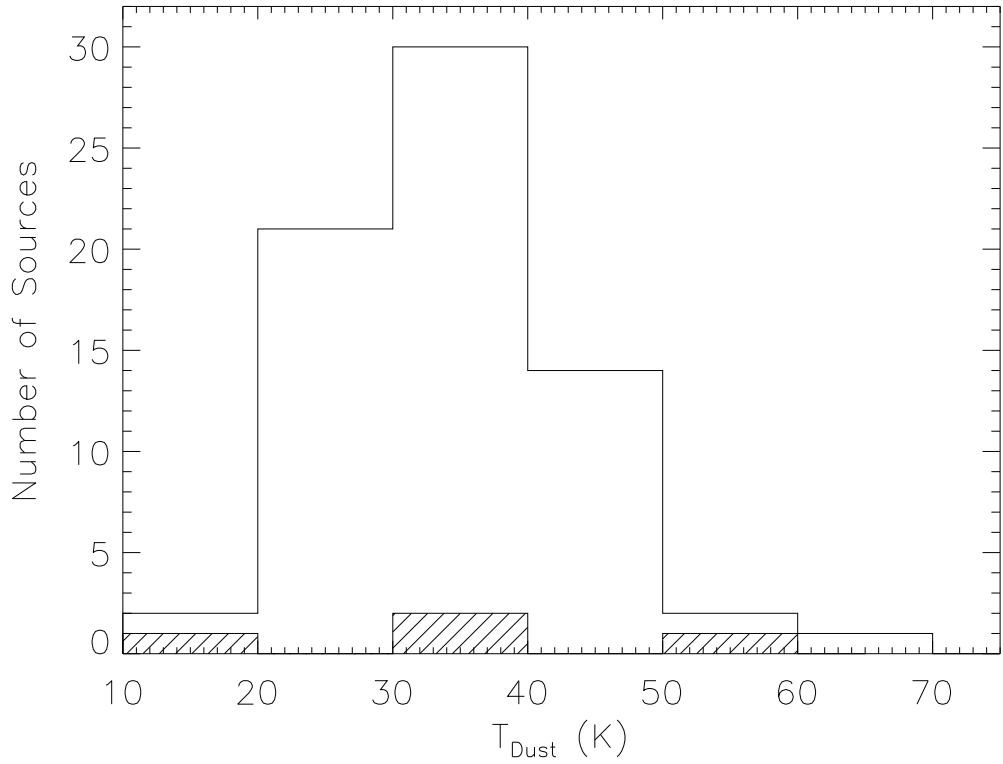}
\caption{Histogram of dust masses, temperatures estimated from single temperature modified black-body fits with an emissivity spectral index $\beta$ of 1.91 (see section 3.4). The dashed histograms shows the result of poor fits with only 2 or 3 photometric points, or 4 points with large errors. \label{mbfr}}
\end{figure*}

\begin{figure*}[!h]
\includegraphics[angle=0,width=0.85\linewidth]{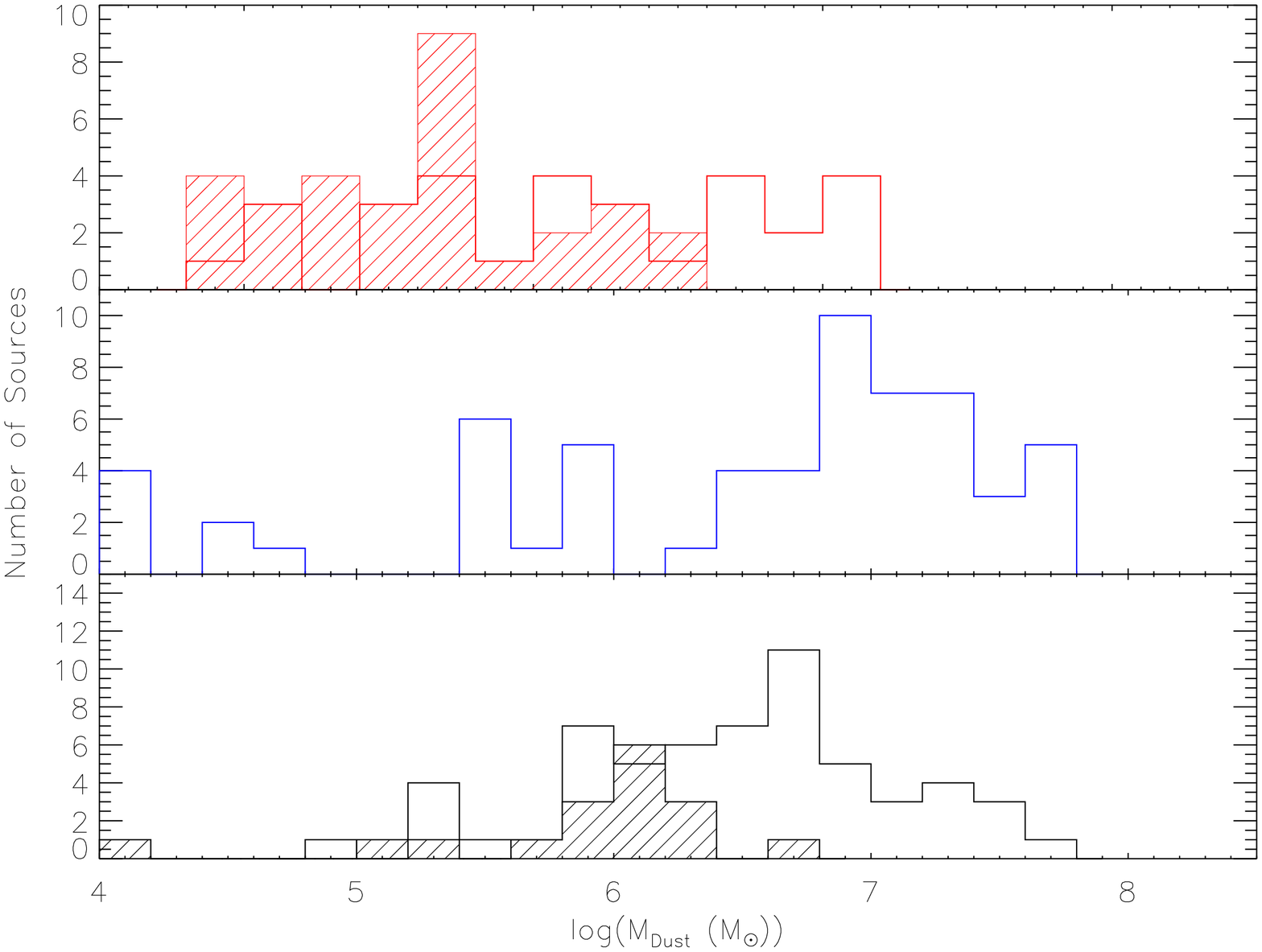}%Mdust_histo_comp666.eps}
\caption{Histograms of dust masses in massive elliptical galaxies (top red), nearby spirals (middle blue) and the PG QSOs presented in this paper (bottom black).\label{MDC}}
\end{figure*}

\begin{figure*}[!h]
\includegraphics[angle=90,width=0.99\linewidth]{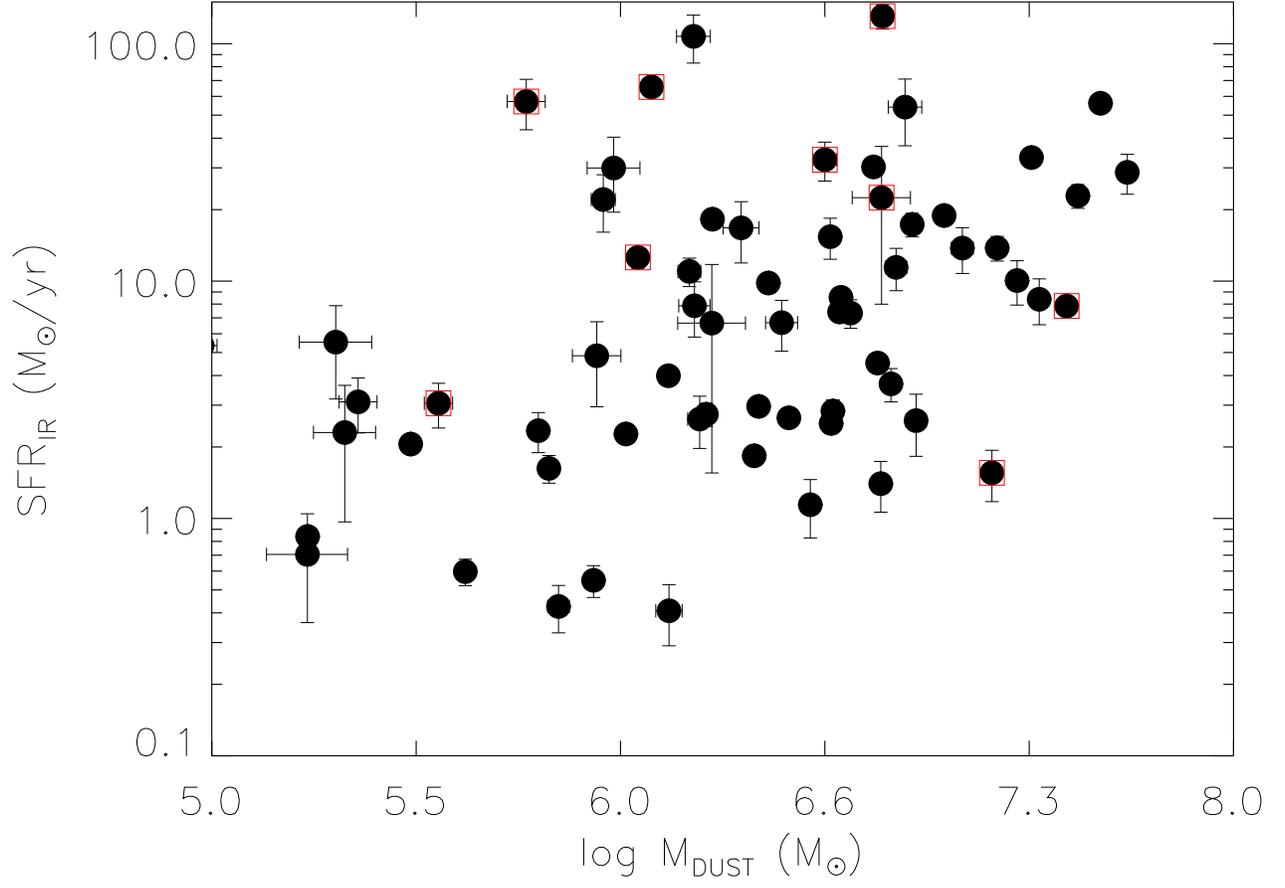}
\caption{Star-formation rates derived from the FIR using the relation of \citep{mur2011} versus total dust masses estimated  from the single temperature modified black body fits.  Red squares represent radio-loud sources, using the \citep{kel1989} definition of radio-loud quasars. \label{SFvMD}}
\end{figure*}

\begin{figure*}[!h]
\includegraphics[angle=90,width=0.99\linewidth]{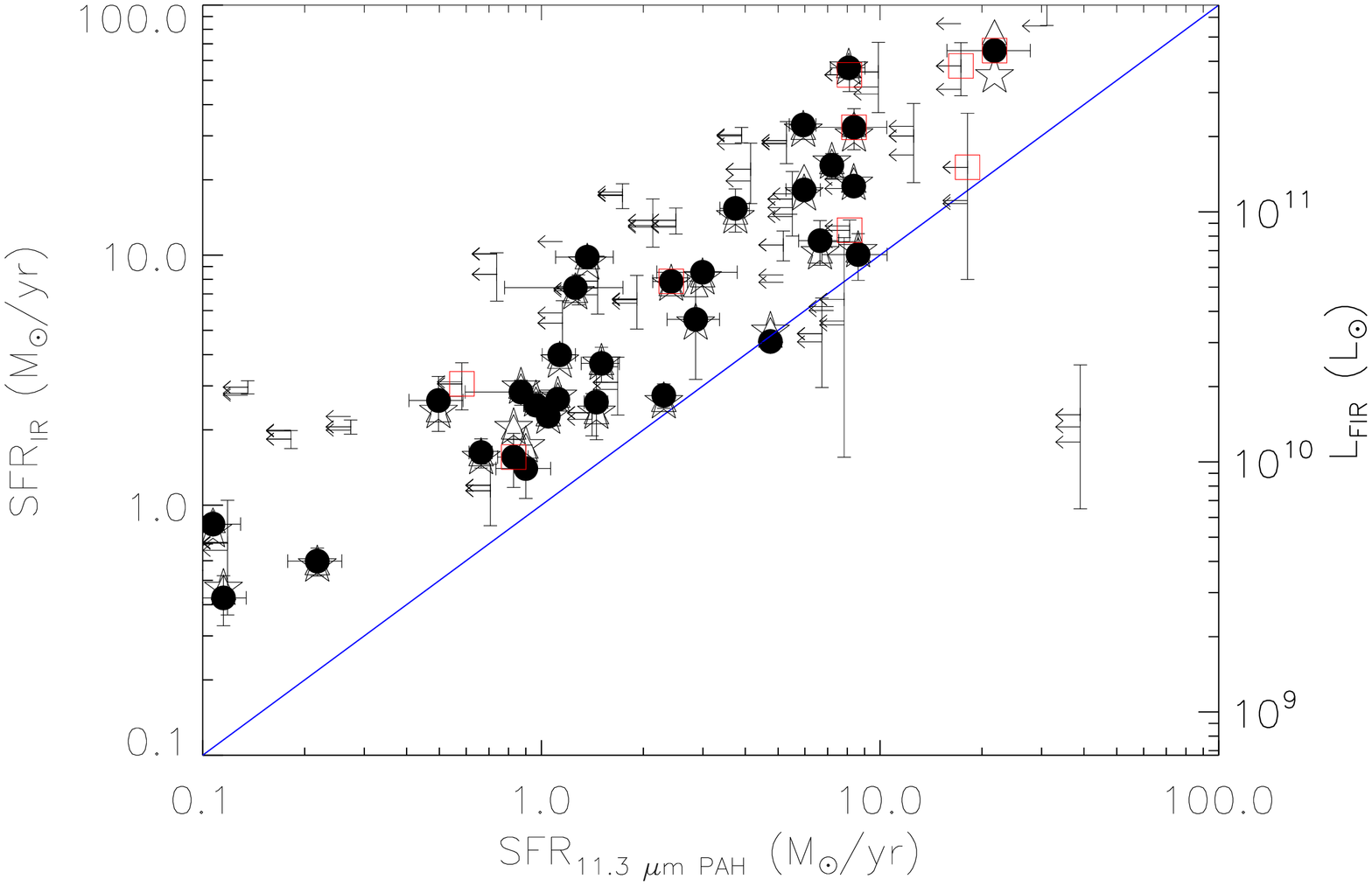}
\caption{\label{spVsf}Star-formation rates derived from the PAH 11.3 $\mu$m measurements of \citep{shi2007} versus the total FIR luminosities, and the SFR derived from the FIR using the relation of \citep{mur2011}. Solid circles are SFR rates estimated from the 42.5 to 122.5$\mu$m FIR as defined by \citet{hel1985}, the stars and triangles are SFR rates estimated from the 40-500 $\mu$m and the 8-1000$\mu$m respectively, as computed from the single temperature modified black body fits. Red squares mark radio-loud sources, using the \citep{kel1989} demarcation of radio-loud quasars.}
\end{figure*}

\begin{figure*}[!h]
$
\begin{array}{cc}
\includegraphics[angle=90,width=0.49\linewidth]{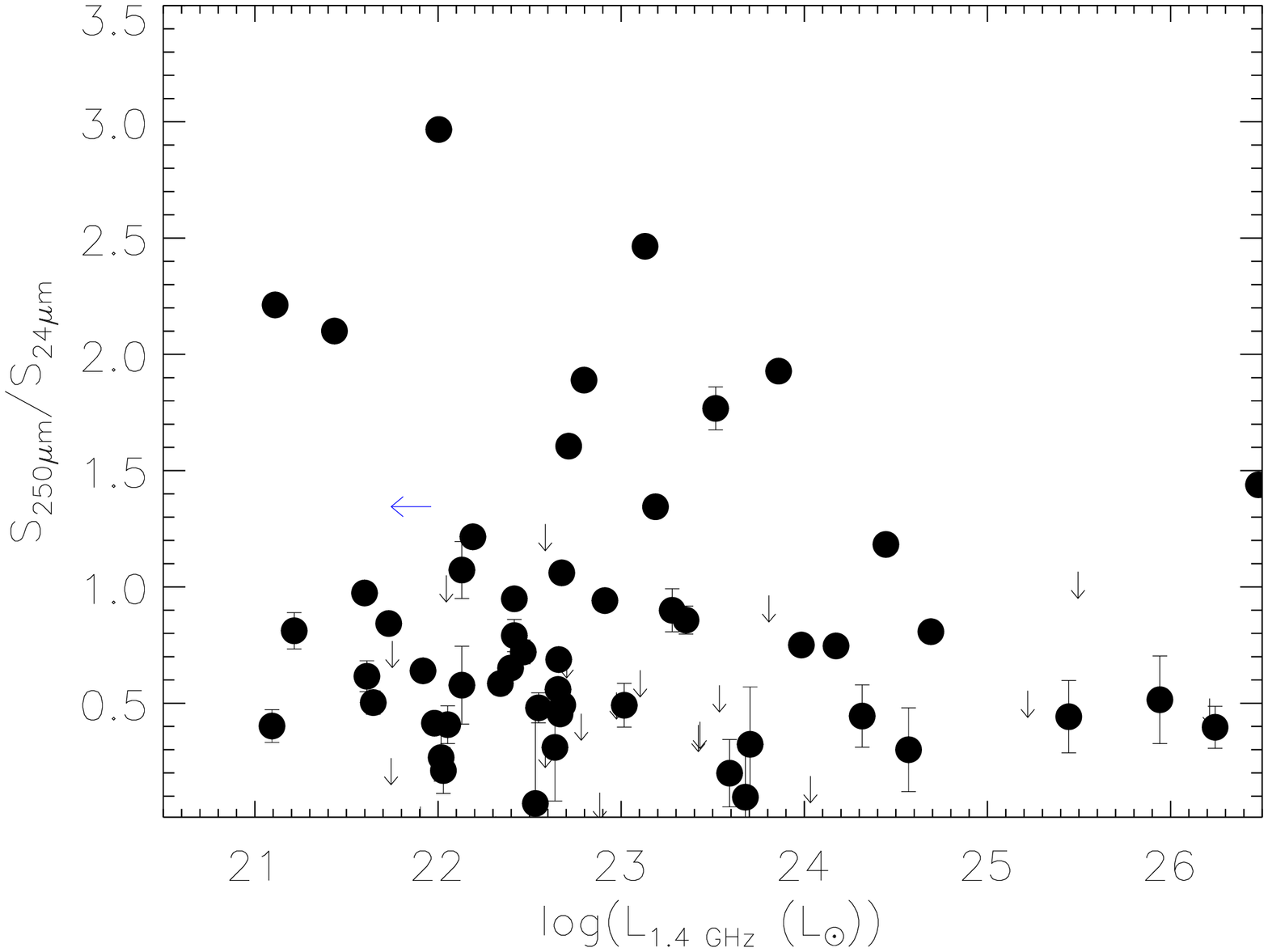}
\includegraphics[angle=90,width=0.49\linewidth]{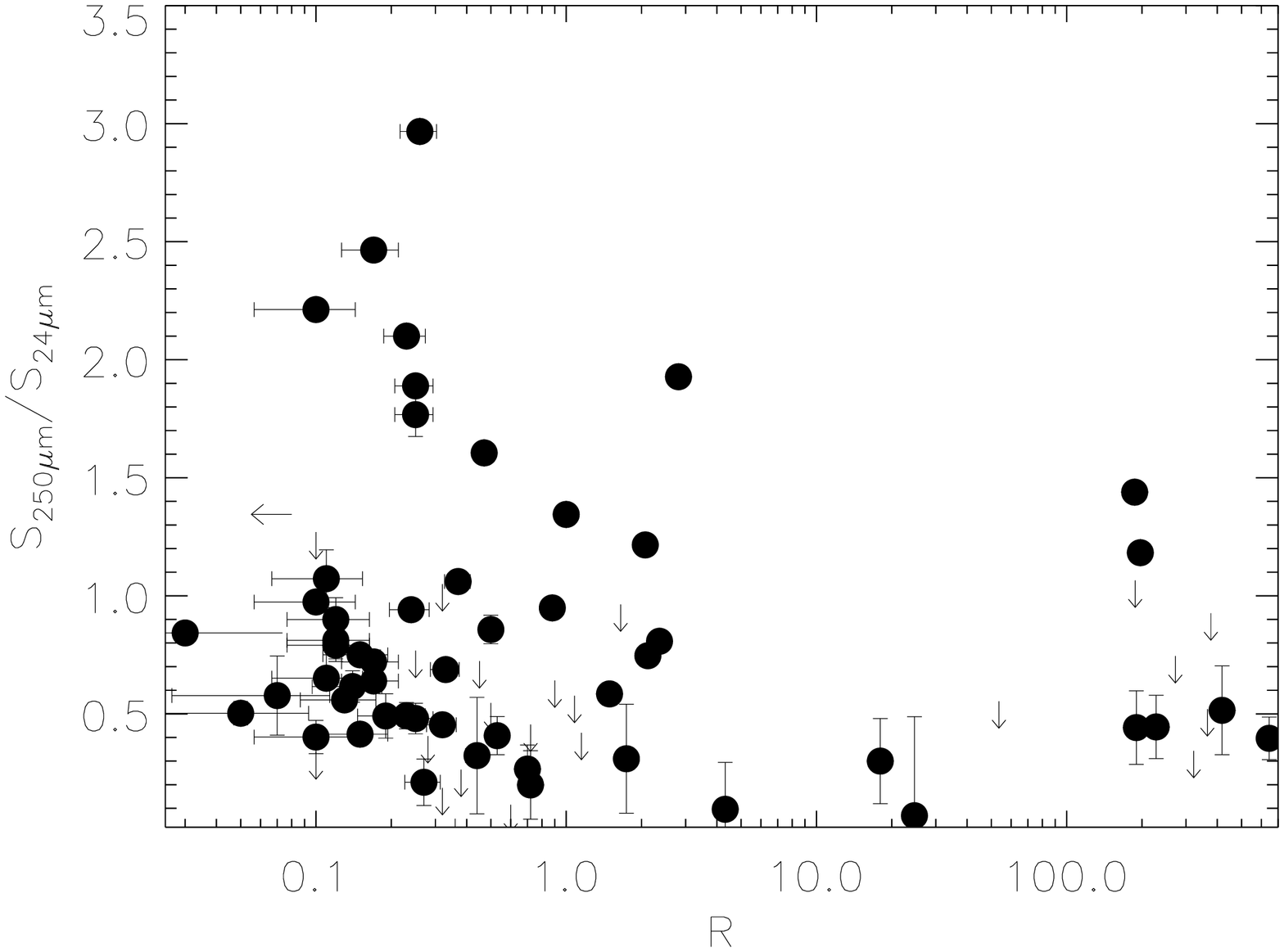}\\
\includegraphics[angle=90,width=0.49\linewidth]{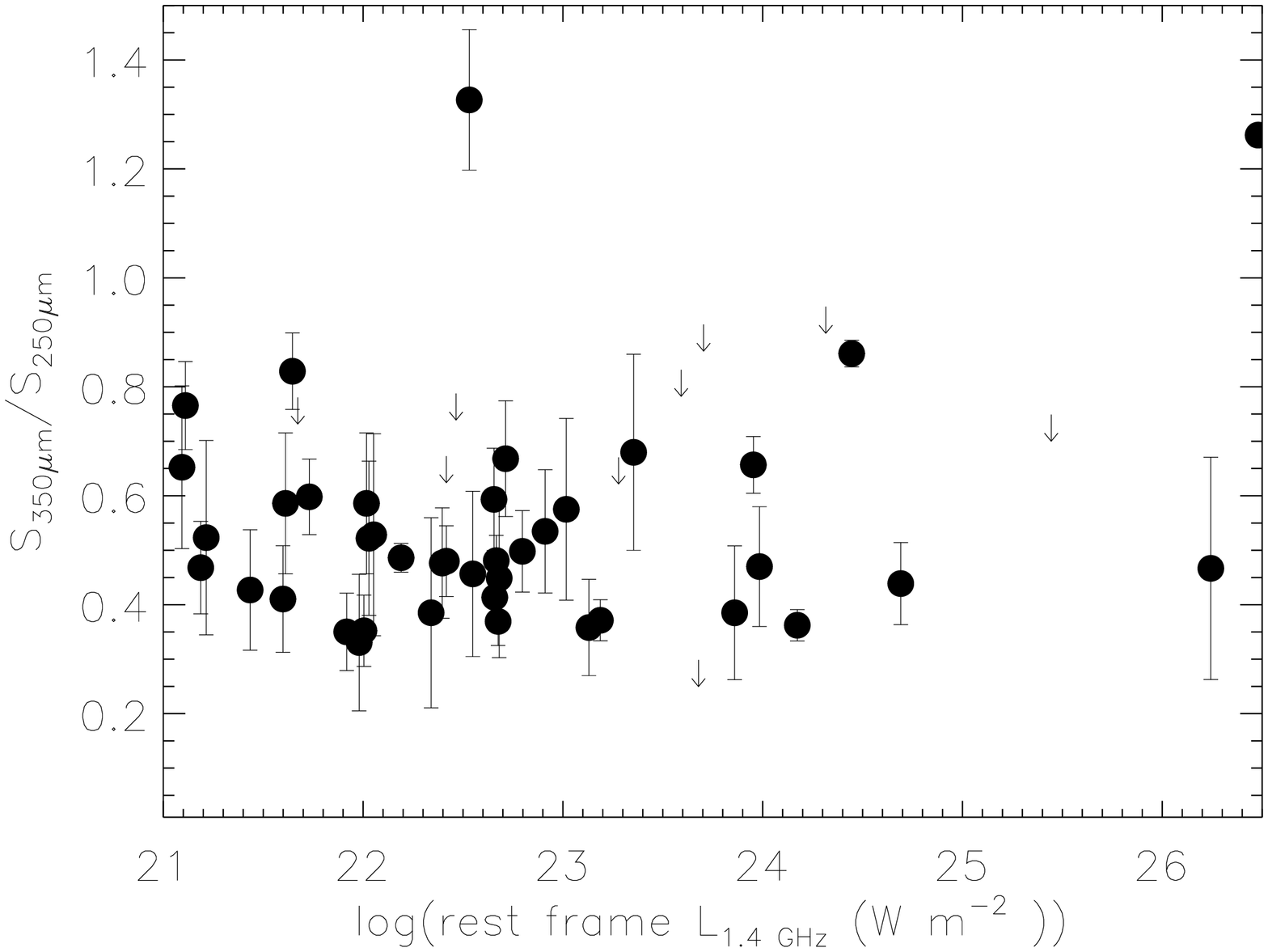}
\includegraphics[angle=90,width=0.49\linewidth]{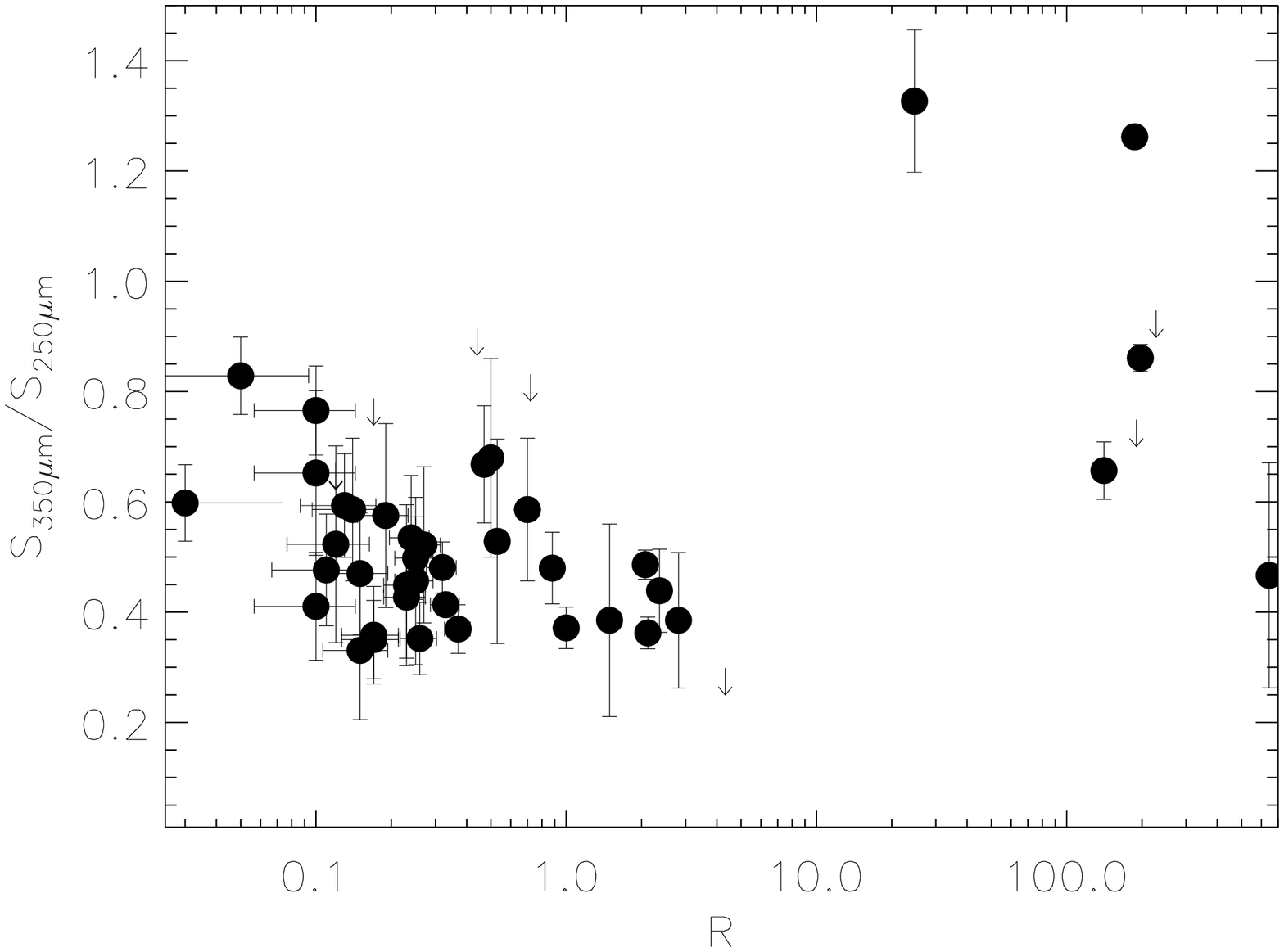}\\
\includegraphics[angle=90,width=0.49\linewidth]{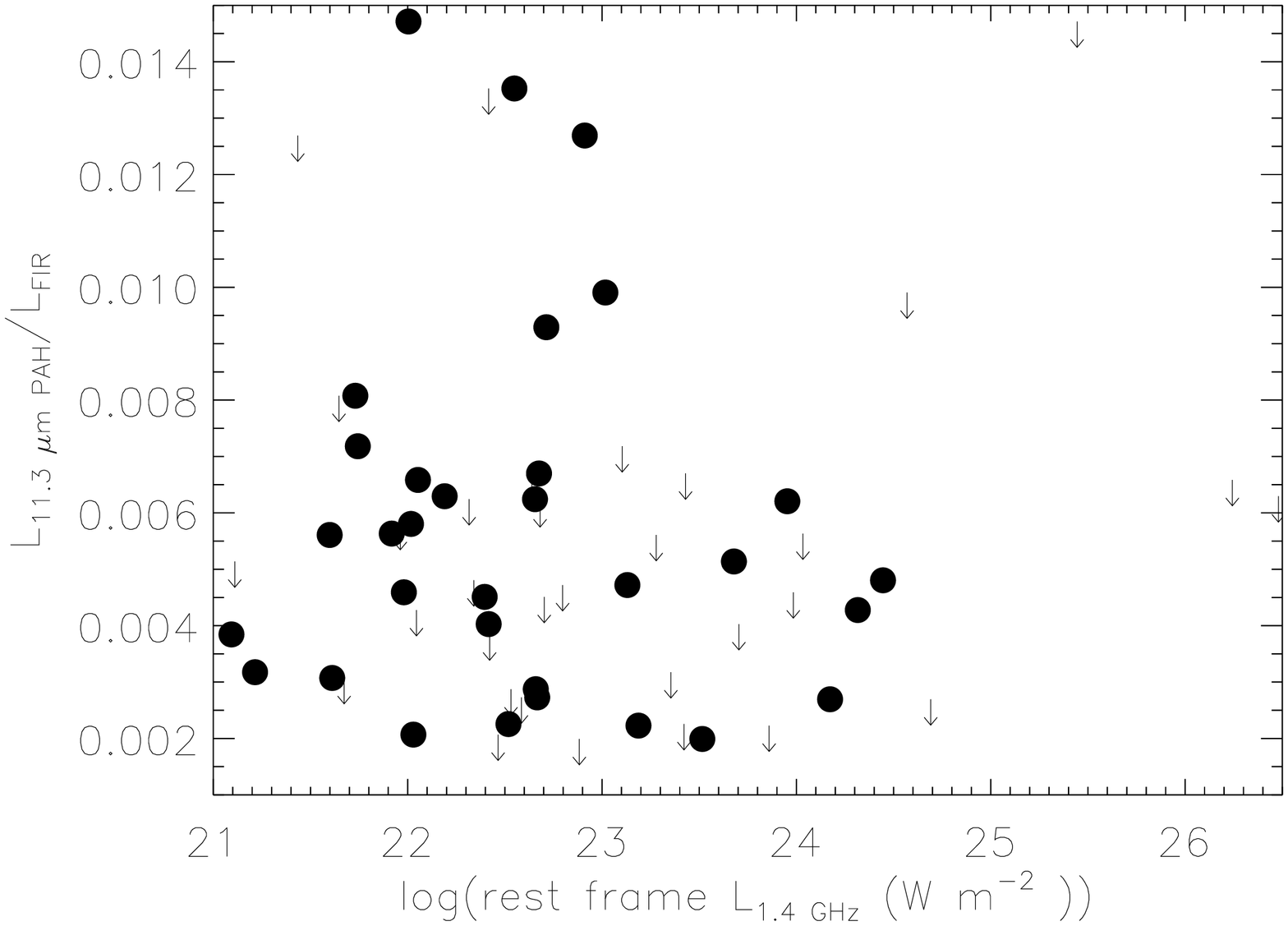}
\includegraphics[angle=90,width=0.49\linewidth]{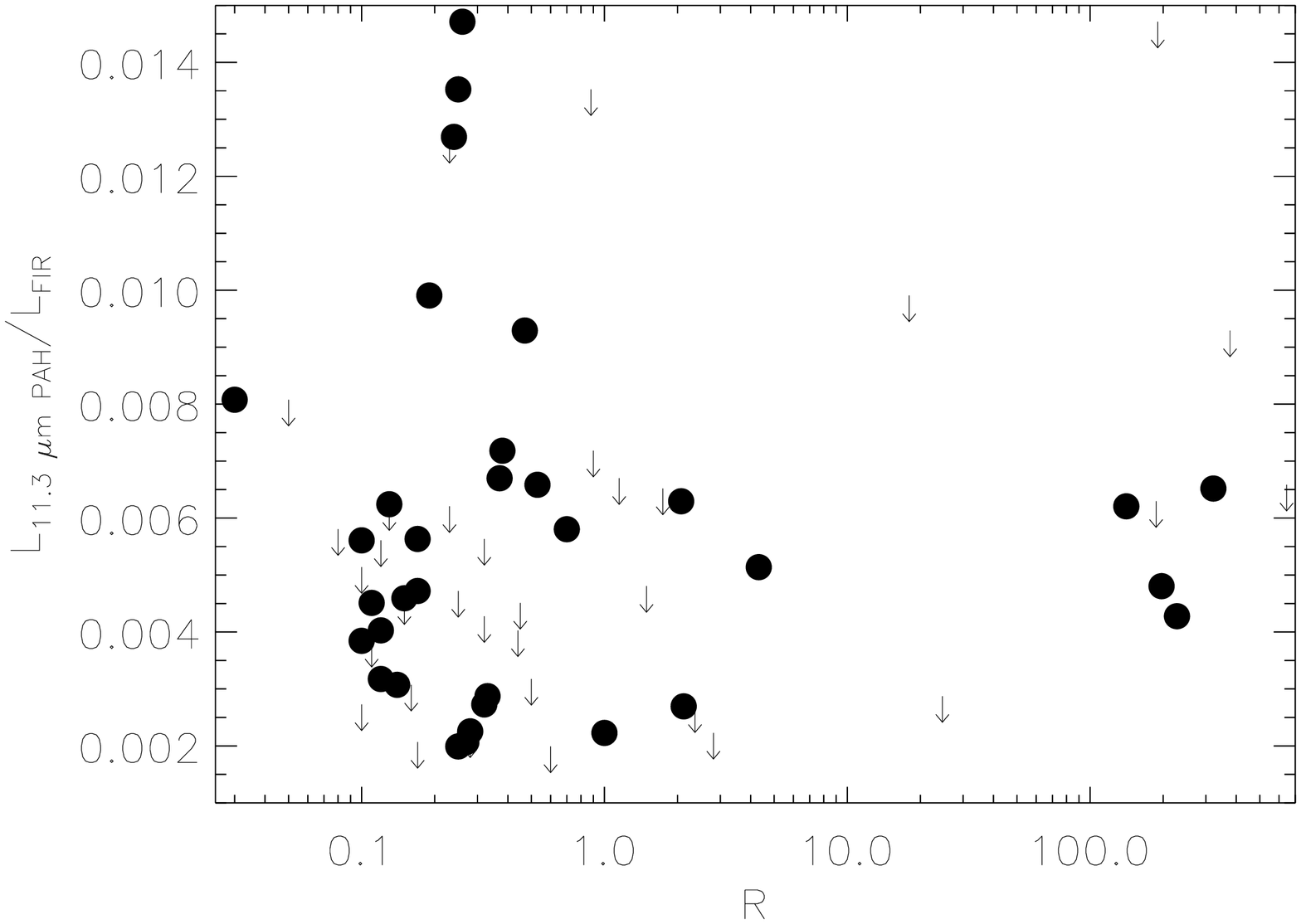}\\
\end{array}$
\caption{Log (Radio Luminosity (W/Hz)) (left) and ratio of radio to optical luminosity (R) versus (right) versus: (top) the ratio of 250 $\mu$m rest-frame Luminosity to the 24 $\mu$m rest-frame  luminosity, (middle) the ratio of 350 $\mu$m rest-frame Luminosity to the 250 $\mu$m rest-frame  luminosity, (bottom) the ratio of 11.3 $\mu$m PAH luminosity to the FIR luminosity. \label{LradVdust}}
\end{figure*}

\clearpage
\bibliographystyle{apj}
\bibliography{../BIBLIO/AP_bib}

\LongTables

\pagestyle{empty}

\clearpage
\begin{deluxetable*}{l c c c c c c c c c c c c}
\tableheadfrac{0.0}
\tablecolumns{13}  
\tabletypesize{\small}    
\tablewidth{0pt}
\tablecaption{FIR Fluxes from PACS  and SPIRE photometry \label{Pphot}}
\tablehead{
\colhead{Name} 
&\colhead{F$_{70\mu m}$ } 
&\colhead{$\sigma _{70\mu m}$ }  
&\colhead{F$_{100\mu m}$ } 
&\colhead{$\sigma _{100\mu m}$ }  
&\colhead{F$_{160\mu m}$ } 
&\colhead{$\sigma _{160\mu m}$ }  
&\colhead{F$_{250\mu m}$ } 
&\colhead{$\sigma _{250\mu m}$ }  
&\colhead{F$_{350\mu m}$ } 
&\colhead{$\sigma _{350\mu m}$ }  
&\colhead{F$_{500\mu m}$ } 
&\colhead{$\sigma _{500\mu m}$ }  \\
\colhead{} 
&\colhead{[mJy]}
&\colhead{[mJy]}
&\colhead{[mJy]}
&\colhead{[mJy]}
&\colhead{[mJy]}
&\colhead{[mJy]}
&\colhead{[mJy]}
&\colhead{[mJy]}
&\colhead{[mJy]}
&\colhead{[mJy]}
&\colhead{[mJy]}
&\colhead{[mJy]}\\
}
\startdata
PG 0003+158      &          28     &         4     &          23     &        13.   & $<$         24     &       ... 	  & $<$          30     &      ... & $<$              25  &      ...  &     $<$ 24 &  ...   \\ 
PG 0003+199      &         341     &        13     &         235     &        17     &         112     &        38       &          66      &         7      &          34     &         7.   & $<$         23     &       ...   \\ 
PG 0007+106      &         234     &        13     &         258     &        13     &         235     &        51   	    &         181      &         7      &         156      &         6.      &         154      &         9.     \\ 
PG 0026+129      &          24     &        14     &          25     &         8.   & $<$         25     &        ...	  & $<$         25.      &     ...   & $<$         22      &    ...   & $<$         21      &       ...    \\ 
PG 0043+039      &          61     &        13     &          55     &        13     &          78     &        31  	    &          31      &         6.   & $<$         23      &       ...  & $<$         22      &        ...     \\ 
PG 0049+171      &          45     &        16     &          39     &        13     &          32     &        18    & $<$         23      &        ...   & $<$         26      &     ...   & $<$         23      &      ...     \\ 
PG 0050+124      &        2237     &        18     &        2436     &        17     &        1861     &        38   &         724     &         7      &         299      &         6      &         128      &         8     \\ 
PG 0052+251      &          78     &        14     &         116     &        13     &         129     &        35  	   &          77      &         7      &          41      &         7   & $<$         22      &       ...     \\
\enddata
\end{deluxetable*}
\clearpage

\begin{deluxetable}{lcccccccc}
%\begin{sidewaystable}
%\rotate
\tableheadfrac{0.0}
\tablecolumns{9}  
\tabletypesize{\small}    
%\tablewidth[{0pt}]
\tablewidth{0pt}
\tablecaption{FIR Photometry Statistics \label{PhotStats}}
\tablehead{
\colhead{Data}
&\colhead{Detection}
&\colhead{Minimum}
&\colhead{Maximum}
&\colhead{Median }
&\colhead{Mean }
&\colhead{Standard}
&\colhead{Median}
&\colhead{Median Upper }\\
\colhead{}
&\colhead{rate}
&\colhead{[mJy] }
&\colhead{[mJy]}
&\colhead{[mJy] }
&\colhead{[mJy]}
&\colhead{Deviation}
&\colhead{Error [mJy] }
&\colhead{Limit [mJy]}\\
%&\colhead{Median $\nu \rm{L}_{\nu obs} [10^9 \rm{L}_{\odot}$]}
%&\colhead{Percentage of light within aperture}
}
\startdata
 PACS  70$\mu$m     &       98\%      &        14    &      2334     &        94    &       212     &       370     &        13      &        53   \\ 
 PACS  100$\mu$m      &       95\%      &         9      &      2436      &       107      &       226     &       382      &        13      &        56  \\ 
 PACS  160$\mu$m   &       81\%      &        15     &      1861      &       113      &       209      &       292      &        38      &       174   \\ 
 SPIRE   250$\mu$m    &       69\%      &        14      &       724      &        63      &       104      &       121     &         6      &        59   \\ 
 SPIRE   350$\mu$m       &       55\%     &        12      &       299      &        41      &        61      &        55      &         6      &        50  \\ 
 SPIRE  500$\mu$m       &       25\%     &        15      &       277      &        34      &        58      &        63      &         7      &        55 
\enddata
%\end{sidewaystable}
\end{deluxetable}
%\end{turnpage}
\clearpage

\begin{landscape}
\begin{deluxetable*}{l r c c c c c c c c c c c c c c c c c}
%\begin{sidewaystable}
\tableheadfrac{0.0}
\tablecolumns{19}  
\tabletypesize{\tiny}    
\tablewidth{0pt}
\tablecaption{NIR, MIR, and Radio Fluxes \label{NMIRtab}}
\tablehead{
\colhead{Name}&\colhead{z}&\colhead{J}&\colhead{$\sigma$ J}&\colhead{H}&\colhead{$\sigma$ H}&\colhead{Ks}&\colhead{$\sigma$ Ks }&\colhead{F$_{3.4\mu m}$}&\colhead{$\sigma$ }&\colhead{F$_{4.6\mu m}$}&\colhead{$\sigma$ }&\colhead{F$_{12\mu m}$}&\colhead{$\sigma$ }&\colhead{F$_{22\mu m}$}&\colhead{$\sigma$ }&\colhead{L$_{1.4 GHz}$}&\colhead{Ref. \tablenotemark{a}}&\colhead{R\tablenotemark{b}} \\
\colhead{}
&\colhead{}
&\colhead{ [mJy] }
&\colhead{[$\mu$Jy]}
&\colhead{[mJy]}
&\colhead{[$\mu$Jy]}
&\colhead{[mJy]}
&\colhead{[$\mu$Jy] }
&\colhead{ [mJy] }
&\colhead{ [$\mu$Jy] }
&\colhead{ [mJy] }
&\colhead{ [$\mu$Jy] }
&\colhead{ [mJy] }
&\colhead{ [$\mu$Jy] }
&\colhead{ [mJy] }
&\colhead{ [mJy] }
&\colhead{Log [W/m$^2$]}
&\colhead{}
&\colhead{}
}
\startdata                  
PG 0003+158    &  0.451    &      2.0    &      5    &      2.1    &      6    &      2.6    &      9    &     3.93    &     10    &     5.86    &     10    &    11.75    &     30    &    28    &      1    &      26.68    &                 2    &   175.00   \\ 
PG 0003+199    &  0.026    &     14.3    &     34    &     20.8    &     74    &     38.7    &     88    &    73.78    &    180    &   100.41    &    250    &   163.19    &    400    &   310    &      2    &      22.03    &                 2    &     0.27   \\ 
PG 0007+106    &  0.089    &      4.5    &     13    &      6.3    &     23    &     12.9    &     26    &    22.49    &     50    &    29.28    &     60    &    55.71    &    130    &   142  &      1    &      24.45    &                 3     &   197.00   \\ 
PG 0026+129    &  0.142    &      5.9    &     13    &      6.1    &     19    &      9.5    &     19    &    13.02    &     30    &    17.58    &     40    &    27.36    &    120    &    49    &      1    &      23.54    &                 2    &     1.08   \\ 
PG 0043+039    &  0.385    &      1.9    &      4    &      1.9    &      7    &      2.4    &      6    &     5.24    &     10    &     7.65    &     20    &    13.57    &     50    &    25    &      1    &      23.28    &  1 &     0.12   \\ 
PG 0049+171    &  0.064    &      1.8    &      5    &      2.0    &      8    &      3.1    &      9    &     5.46    &     10    &     7.39    &     10    &    13.39    &     30    &    20    &      1    &      22.05    &  1 &     0.32   \\ 
PG 0050+124    &  0.061    &     16.6    &     40    &     26.0    &     90    &     48.1    &    107    &    88.29    &    240    &   130.07    &    300    &   389.68    &   1140    &   944   &      8    &      22.66    &                 3     &     0.33   \\
\enddata
\tablenotetext{a}{1: FIRST survey \citep{bwh1995}, 2:  NVSS survey \citep{condon1992}, 3: data published in \citet{kel1989}}
\tablenotetext{b}{Radio Loudness as described in section 3.5, introduced and measured by \citet{kel1989}.}
\end{deluxetable*}
\clearpage
\vfill
\clearpage
\newpage
\end{landscape}

\begin{landscape}
\begin{deluxetable*}{lcccccccccccc}
\tableheadfrac{0.0}
\tablecolumns{13}  
\tabletypesize{\tiny}    
\tablewidth{0pt}
\tablecaption{Rest Frame Luminosities from MIR and FIR photometry with Herschel and WISE \label{LumRestTab} }
\tablehead{
\colhead{Name}
&\colhead{$\nu$L$_{\nu 5.5 \mu \rm{m}}$}
&\colhead{$\nu$L$_{\nu 8 \mu \rm{m}}$}
&\colhead{$\nu$L$_{\nu 24\mu \rm{m}}$}
&\colhead{$\nu$L$_{\nu 60\mu \rm{m}}$}
&\colhead{$\nu$L$_{\nu 70\mu \rm{m}}$}
&\colhead{$\nu$L$_{\nu 100\mu \rm{m}}$}
&\colhead{$\nu$L$_{\nu 160\mu \rm{m}}$}
&\colhead{$\nu$L$_{\nu 250\mu \rm{m}}$}
&\colhead{$\nu$L$_{\nu 350\mu \rm{m}}$}
&\colhead{$\nu$L$_{\nu 500\mu \rm{m}}$}
&\colhead{L$_{FIR1}$\tablenotemark{a}}
%&\colhead{L$_{IR2}$\tablenotemark{b}}
&\colhead{L$_{FIR}$\tablenotemark{b}}\\
%&\colhead{Percentage of light within aperture}
\colhead{}
&\colhead{Log(L$_{\odot}$)}
&\colhead{Log(L$_{\odot}$)}
&\colhead{Log(L$_{\odot}$)}
&\colhead{Log(L$_{\odot}$)}
&\colhead{Log(L$_{\odot}$)}
&\colhead{Log(L$_{\odot}$)}
&\colhead{Log(L$_{\odot}$)}
&\colhead{Log(L$_{\odot}$)}
&\colhead{Log(L$_{\odot}$)}
&\colhead{Log(L$_{\odot}$)}
&\colhead{Log(L$_{\odot}$)}
%&\colhead{Log(L$_{\odot}$)}
&\colhead{Log(L$_{\odot}$)}\\
}
\startdata
PG 0003+158  &     11.9  &     11.9  &     11.8  & 11.4  & 11.3  &$<$12.6  & $<$  10.9  &$<$12.2  &$<$12.0  &$<$11.9  & $<$  12.1  &       ... \\ 
PG 0003+199  &     10.4  &     10.2  &     10.2  & 9.8  & 9.7  & 9.4  & 8.9  & 8.5  & 8.1  & $<$   7.8  & 9.8  &         9.7 \\ 
PG 0007+106  &     11.0  &     10.9  &     11.0  & 10.7  & 10.7  & 10.6  & 10.3  & 10.0  & 9.8  & 9.7  & 10.7  &        10.7 \\ 
PG 0026+129  &     11.2  &     11.1  &     10.9  & 10.2  & 10.1  &$<$11.4  & $<$   9.8  &$<$11.0  &$<$10.9  &$<$10.7  & $<$  11.0  &       ... \\ 
PG 0043+039  &     11.9  &     11.8  &     11.7  & 11.6  & 11.6  & 11.4  & 11.4  & 10.7  & $<$  10.4  &$<$11.7  & 11.6  &       11.5 \\ 
PG 0049+171  &     10.1  &     10.0  &      9.8  & 9.7  & 9.7  & 9.4  & 9.2  & $<$   8.8  &$<$10.1  &$<$10.0  & 9.7  &         9.6 \\ 
PG 0050+124  &     11.3  &     11.3  &     11.5  & 11.4  & 11.3  & 11.2  & 10.9  & 10.3  & 9.8  & 9.2  & 11.3  &        11.3 \\ 
PG 0052+251  &     11.3  &     11.2  &     11.2  & 10.8  & 10.8  & 10.8  & 10.6  & 10.2  & 9.8  & $<$   9.5  & 10.8  &        10.8 \\
\enddata
\tablenotetext{a}{L$_{FIR}$  [42.5-122.5$\mu$m] as defined in \citet{hel1985}}
%\tablenotetext{b}{L$_{IR}$ [8-1000$\mu$] integrated over the best modified black body fit.}
\tablenotetext{b}{L$_{FIR}$ [40-500$\mu$m] integrated over the best modified black body fit.}
\end{deluxetable*}
\clearpage
\vfill
\clearpage
\newpage
\end{landscape}

\begin{deluxetable}{llllll}
\tableheadfrac{0.0}
\tablecolumns{6}  
\tabletypesize{\small}    
\tablewidth{0pt}
\tablecaption{Basic Statistics of IR Luminosities \label{LirStats}}
\tablehead{
\colhead{Restframe}
&\colhead{Minimum Detection}
&\colhead{Maximum Detection}
&\colhead{Median Detection }
&\colhead{Kaplan-Meier Median}
&\colhead{95\% confidence interval}\\
\colhead{Luminosity}
&\colhead{[L$_{\odot}$]}
&\colhead{[L$_{\odot}$]}
&\colhead{[L$_{\odot}$]}
&\colhead{[L$_{\odot}$]}
&\colhead{[L$_{\odot}$]}\\
}
\startdata
L$_{60\mu m}$ & $1.4 ~\times 10^9$ & $2.8 ~\times 10^{13}$ & $6 ~\times 10^{10}$ & $6 ~\times 10^{10}$ & $4-10 ~\times 10^{10}$ \\
L$_{100\mu m}$ & $1.5 ~\times 10^9$ & $1.0~\times 10^{12}$ & $2 ~\times 10^{10}$ & $3 ~\times 10^{10}$ & $2-5 ~\times 10^{10}$  \\
L$_{160\mu m}$ & $5~\times 10^7$ & $4.5 ~\times 10^{11}$ & $2 ~\times 10^{10}$ & $1 ~\times 10^{10}$ & $0.8-2 ~\times 10^{10}$  \\
L$_{250\mu m}$ & $3~\times 10^8$ & $2.6 ~\times 10^{11}$ & $6 ~\times 10^{9}$ & $4 ~\times 10^{9}$ & $2-9 ~\times 10^{9}$  \\
L$_{350\mu m}$ & $1 ~\times 10^8$ & $1.1 ~\times 10^{11}$ & $3 ~\times 10^{9}$ &$1 ~\times 10^{9}$ & $0.9-3 ~\times 10^{9}$  \\
L$_{500\mu m}$ & $2~\times 10^8$ & $1.1 ~\times 10^{11}$ & $9 ~\times 10^{9}$ & $4 ~\times 10^{8}$ & $2-7 ~\times 10^{8}$  \\
\enddata
\end{deluxetable}

\clearpage
%\begin{turnpage}
\begin{deluxetable}{ccc}
%\begin{sidewaystable}
%\rotate
\tableheadfrac{0.0}
\tablecolumns{3}  
%\tabletypesize{\small}    
%\tablewidth[{0pt}
\tablecaption{Normalized Templates at 6 $\mu$m \label{TempNorm}}
\tablehead{
\colhead{$\lambda _{\rm{rest}}$} 
&\colhead{F$_{median}$}
&\colhead{95\% Interval }\\
}
\startdata
%\hline
%\multicolumn{3}{r}{Template of fluxes normalized at 5 $\mu$m rest-frame.} \\
%\hline
%1.2 & 0.15 & 0.14 - 0.17 \\
%1.6 & 0.19 & 0.18 - 0.21 \\
%1.5 & 0.28 & 0.26 - 0.30 \\
%3.1 & 0.57 & 0.52 - 0.59 \\
%3.5 & 0.75 & 0.73 - 0.77 \\
%10.7 & 1.62 & 1.56 - 1.74 \\
%21.0 & 3.45 & 3.17 - 3.99 \\
%51.0 & 5.17 & 3.56 - 6.38 \\
%86.6 & 5.25 & 4.04 - 6.42 \\
%137.0 & 4.09 & 2.37 - 6.20\\
%214.9 & 1.74 & 1.33 - 2.81\\
%322.0 & 1.03 & 0.79 - 1.61\\
%429.2 & 0.44 & 0.27 - 0.64\\
1.1 & 0.15 & 0.13 - 0.17 \\
1.5 & 0.21 & 0.19 - 0.23 \\
2.0 & 0.32 & 0.28 - 0.34 \\
3.0 & 0.55 & 0.51 - 0.58 \\
4.1 & 0.73 & 0.70 - 0.75 \\
10.4 & 1.51 & 1.45 - 1.61 \\
19.0 & 3.20 & 2.97 - 3.74 \\
60.8 & 4.63 & 3.26 - 5.88 \\
87.2 & 5.06 & 3.61 - 6.52 \\
138.1 & 3.76 & 2.05 - 5.71\\
217.6 & 1.60 & 0.90 - 2.31\\
308.4 & 0.80 & 0.73 - 1.34\\
436.0 & 0.46 & 0.25 - 0.68\\
\enddata
\end{deluxetable}
\clearpage

%\begin{landscape}
\begin{deluxetable*}{lcccccc}
\tableheadfrac{0.0}
\tablecolumns{7}  
%\tabletypesize{\small}    
\tablewidth{0pt}
\tablecaption{Cold Dust Estimates \label{TabR1}}
\tablehead{
\colhead{Name}
&\colhead{log(M1$_D$)\tablenotemark{a}}
&\colhead{T1$_D$ [K]}
&\colhead{Fitting Errors}
&\colhead{log(M2$_D$)\tablenotemark{b}}
&\colhead{T2$_D$ [K]}
&\colhead{Fitting Errors}\\
\colhead{}
&\colhead{[M$_\odot$]}
&\colhead{$\beta=1.91$}
&\colhead{}
&\colhead{[M$_\odot$]}
&\colhead{$\beta=2.08$}
&\colhead{}\\
}
\startdata
%2p          PG 0003+199  &      5.2  &         39  &   3\%  &      5.3  &         37  &   3\% \\ 
 %NT        PG 0007+106  &      7.4  &         23  &   2\%  &      7.6  &         22  &   2\% \\ 
          PG 0043+039  &      6.9  &         36  &  12\%  &      7.0  &         35  &  12\% \\ 
          PG 0049+171  &      5.2  &         37  &  22\%  &      5.3  &         36  &  22\% \\ 
          PG 0050+124  &      7.3  &         32  &   0\%  &      7.4  &         30  &   0\% \\ 
          PG 0052+251  &      7.2  &         26  &   6\%  &      7.4  &         25  &   6\% \\ 
          PG 0157+001  &      7.6  &         40  &   1\%  &      7.7  &         38  &   1\% \\ 
          PG 0804+761  &      5.0  &         66  &   8\%  &      5.1  &         61  &   8\% \\ 
          PG 0838+770  &      7.2  &         28  &   3\%  &      7.3  &         26  &   3\% \\ 
          PG 0844+349  &      6.8  &         23  &   5\%  &      6.9  &         22  &   5\% \\
\enddata
  \tablenotetext{a}{Values determined from modified blackbody fits, with absorption coefficients and spectral indices derived from the dust model of \citep{comp2011}.}
  \tablenotetext{b}{Values determined from modified blackbody fits, with absorption coefficients and spectral indices derived from the dust model of \citep{dli2007}}

\end{deluxetable*}
\clearpage
%\end{landscape}

\clearpage

% \begin{landscape}

%\end{landscape}

\end{document}